\documentclass[12pt]{article}
\pdfoutput=1
\usepackage[authoryear]{natbib}
\usepackage{authblk}
\usepackage{array}
\usepackage{multirow}

\usepackage{amsbsy,amsmath,amsthm,latexsym,amssymb}
\usepackage{graphicx,stmaryrd,sectsty}
\usepackage{setspace}
\usepackage[font=small,labelfont=bf]{caption}
\usepackage{verbatim}
\usepackage{graphicx}
\usepackage{url}
\newcommand{\bc}{\begin{center}}
	\newcommand{\ec}{\end{center}}
\newcommand{\bfr}{\begin{flushright}}
	\newcommand{\efr}{\end{flushright}}

\newcommand{\no}{\noindent}
\newcommand{\be}{\begin{enumerate}}
	\newcommand{\ee}{\end{enumerate}}
\newcommand{\bi}{\begin{itemize}}
	\newcommand{\ei}{\end{itemize}}
\newcommand{\bd}{\begin{description}}
	\newcommand{\ed}{\end{description}}
\newcommand{\beq}{\begin{equation}}
	\newcommand{\eeq}{\end{equation}}
\newcommand{\bea}{\begin{eqnarray}}
	\newcommand{\eea}{\end{eqnarray}}

\newcommand{\bfi}{\begin{figure}}
	\newcommand{\efi}{\end{figure}}
\newcommand{\bay}{\begin{array}{l}}
	\newcommand{\eay}{\end{array}}

\newcommand{\cref}[1]{(\ref{#1})}   %to make cross reference easy.

\sectionfont{\normalsize}
\subsectionfont{\normalsize}
\subsubsectionfont{\normalsize}
\graphicspath{{./figures/}}

\usepackage{siunitx,stmaryrd}
\usepackage{breqn}
\usepackage{subfig}
\usepackage{float}
\usepackage{xcolor}
\begin{document}
	%\begin{frontmatter}
	
	
	\begin{titlepage}
		\clearpage\thispagestyle{empty}
		%\noindent {\footnotesize {{\em
		%\hfill To be submitted to Cement and Concrete Composites} }} \\
		\noindent
		\hrulefill
		\begin{figure}[h!]
			\centering
			\includegraphics[width=2 in]{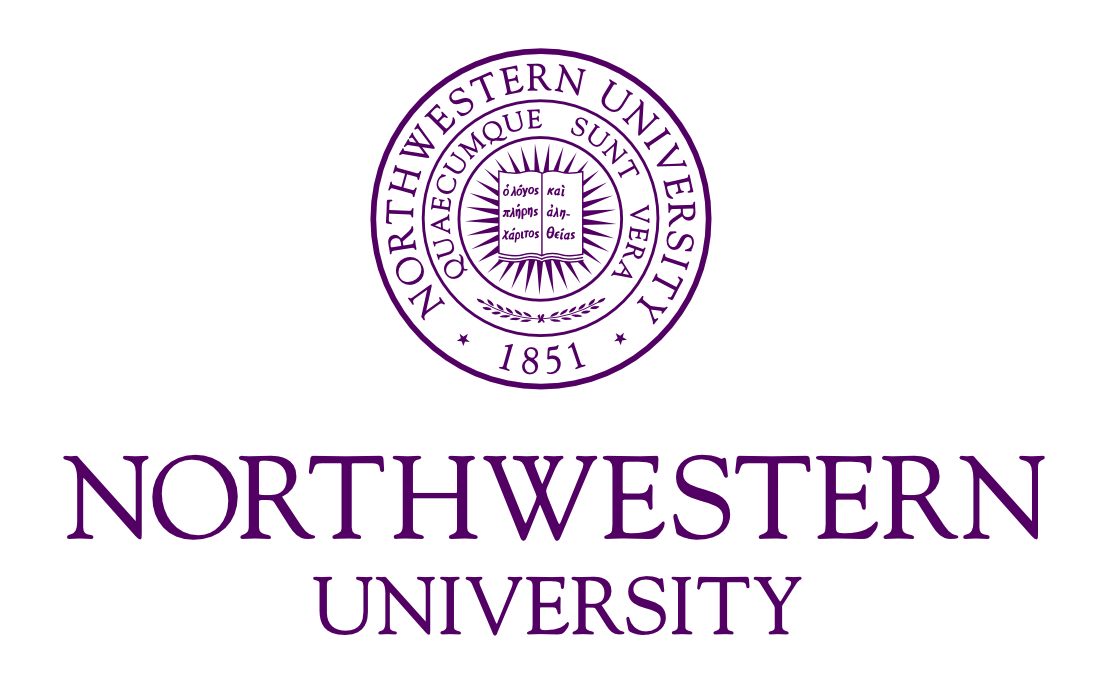}
		\end{figure}
		%{\color{NU} some text}
		\begin{center}
			{%\color{NU}
				{
					{\bf Center for Sustainable Engineering of Geological and Infrastructure Materials} \\ [0.1in]
					Department of Civil and Environmental Engineering \\ [0.1in]
					McCormick School of Engineering and Applied Science \\ [0.1in]
					Evanston, Illinois 60208, USA
				}
			}
		\end{center} %\vskip 5mm
		\hrulefill \\ \vskip 2mm
		\vskip 0.5in
		\begin{center}
			{\large {\bf Multiphysics Lattice Discrete Particle Modeling (M-LDPM) for the Simulation of Shale Fracture Permeability
				}}\\[0.5in]
				{\large {\sc Weixin Li, Xinwei Zhou, J. William Carey, Luke P. Frash, Gianluca Cusatis}}\\[0.75in]
				{\sf \bf SEGIM INTERNAL REPORT No. 18-3/976M}\\[0.75in]
			\end{center}
			\vskip 5mm
			\noindent {\footnotesize {{\em Submitted to Rock Mechanics and Rock Engineering \hfill March 2018} }}
		\end{titlepage}
		
		\newpage
		\clearpage \pagestyle{plain} \setcounter{page}{1}
		%\end{frontmatter}
		%\runningheads{W. Li, R. Rezakhani, C. Jin and G. Cusatis}{A multiscale discrete approach to anisotropic mechanical behavior of shale }
		\title{Multiphysics Lattice Discrete Particle Modeling (M-LDPM) for the Simulation of Shale Fracture Permeability}
		%\author{Weixin Li$_1$, Roozbeh Rezakhani$_2$, Congrui Jin$_3$, Xinwei Zhou$_4$, and Gianluca Cusatis$_{1,2}$}
		%\author{Weixin Li\footnote{
		%	Theoretical and Applied Mechanics, Northwestern University, Evanston, IL 60208, U.S.A.},
		%Roozbeh Rezakhani\footnote{
		%	Department of Civil and Environmental Engineering, Northwestern University, Evanston, IL 60208, U.S.A.}, 
		%Congrui Jin\footnote{Department of Mechanical Engineering, State University of New York at Binghamton, Binghamton, NY 13902, U.S.A.}, 
		%Xinwei Zhou\
		%}
		\author[1]{\small Weixin Li\thanks{Email: w.li@u.northwestern.edu}}
		\author[3]{Xinwei Zhou}
		\author[4]{J. William Carey\thanks{Email: bcarey@lanl.gov}}
		\author[4]{Luke P. Frash}
		\author[1,2]{Gianluca Cusatis \thanks{Correspondence to: Gianluca Cusatis. Email:g-cusatis@northwestern.edu}}
		
		%\address[1]{Theoretical and Applied Mechanics, Northwestern University, Evanston, IL 60208, U.S.A.}
		%\address[2]{Department of Civil and Environmental Engineering, Northwestern University, Evanston, IL 60208, U.S.A.}
		%\address[3]{Department of Mechanical Engineering, State University of New York at Binghamton, Binghamton, NY 13902, U.S.A.}
		%\address[4]{Department of Mechanical Engineering, State University of New York at Binghamton, Binghamton, NY 13902, U.S.A.}
		\affil[1]{\footnotesize Theoretical and Applied Mechanics, Northwestern University, Evanston, IL 60208, U.S.A.}
		\affil[2]{\footnotesize Department of Civil and Environmental Engineering, Northwestern University, Evanston, IL 60208, U.S.A.}	
		\affil[3]{\footnotesize ES3, 550 West C St., San Diego, CA 92101, U.S.A.}		
		\affil[4]{\footnotesize Earth \& Environmental Sciences, Los Alamos National Laboratory, Los Alamos, NM 87545, U.S.A.}	

		%\author{Weixin Li\affil{1},
		%	Roozbeh Rezakhani\affil{2}, Congrui Jin\affil{3} and Gianluca Cusatis\affil{1,2}\corrauth}
		%\address{\affilnum{1}Theoretical and Applied Mechanics, Northwestern University, Evanston, IL 60208, U.S.A.\break
		%	\affilnum{2}Department of Civil and Environmental Engineering, Northwestern University, Evanston, IL 60208, U.S.A.\break
		%	\affilnum{3}Department of Mechanical Engineering, State University of New York at Binghamton, Binghamton, NY 13902, U.S.A.}
		%\corraddr{Gianluca Cusatis, Department of Civil and Environmental Engineering, Northwestern University, Evanston, IL 60208, U.S.A. E-mail: g-cusatis@northwestern.edu }
		%\cgs{<Contract/grant sponsor name (no number)>}
		%\cgsn{<Contract/grant sponsor name>}{<number>}
		\date{}
		\maketitle
		%\begin{abstract}
		{\small \no {\bf   Abstract}: 
		A three-dimensional Multiphysics Lattice Discrete Particle Model (M-LDPM) framework is formulated to investigate the fracture permeability behavior of shale. The framework features a dual lattice system mimicking the mesostructure of the material and simulates coupled mechanical and flow behavior. The mechanical lattice model simulates the granular internal structure of shale, and describes heterogeneous deformation by means of discrete compatibility and equilibrium equations. The network of flow lattice elements constitutes a dual graph of the mechanical lattice system. A discrete formulation of mass balance for the flow elements is formulated to model fluid flow along cracks. The overall computational framework is implemented with a mixed explicit-implicit integration scheme and a staggered coupling method that makes use of the dual lattice topology enabling the seamless two-way coupling of the mechanical and flow behaviors. The proposed model is used for the computational analysis of shale fracture permeability behavior by simulating triaxial direct shear tests on Marcellus shale specimens under various confining pressures. The simulated mechanical response is calibrated against the experimental data, and the predicted permeability values are also compared with the experimental measurements. Furthermore, the paper presents the scaling analysis of both the mechanical response and permeability measurements based on simulations performed on geometrically similar specimens with increasing size. The simulated stress strain curves show a significant size effect in the post-peak due to the presence of localized fractures. The scaling analysis of permeability measurements enables prediction of permeability for large specimens by extrapolating the numerical results of small ones.  
			%\end{abstract}
			
		}
		{\small \no {\bf   Keyword}: Fracture permeability; Triaxial direct shear; Hydro-mechanical coupling; Lattice Discrete Particle Model; Dual lattice}
		
		%	\end{frontmatter}

		\pagestyle{plain}\thispagestyle{empty}
			
%\newpage

\section{Introduction}\label{sec:intro}
Gas shale and shale-like rock play a crucial role in several engineering applications such as hydraulic fracturing and CO$_2$ sequestration, requiring a basic understanding and characterization of their material properties including mechanical and hydraulic behavior. Particularly, it is essential to investigate the fracture- and damage-induced permeability evolution of the material for two opposite purposes: (1) utilizing the low permeability of shale in subsurface energy storage; (2) enhancing gas recovery from unconventional reservoirs.  This involves quantifying local damage and fracturing of material and correlating permeability evolution with damage by taking into account the multiscale character of shale \citep{hyman2016understanding,ilgen2017shales,li2016integrated}. 

Previous experimental studies have shown that fracture permeability involves multiple physical processes, and is influenced by various factors such as fracture aperture and roughness \citep{witherspoon1980validity,zimmerman1996hydraulic,yeo1998effect}, chemical precipitation and dissolution \citep{detwiler2010permeability,noiriel2010intense,elkhoury2013dissolution}, and thermal effects \citep{rutqvist2008analysis}. In particular, open and connected cracks have a strong influence on permeability. A large amount of previous work on fracture permeability has been conducted on prefractured (split or sawn) specimens and has shown that rock permeability is strongly related to density, spacing, orientation, width, and length of fractures within specimens \citep{gutierrez2000stress,davy2007permeability,jobmann2010investigation,cho2013pressure}. However, researchers suggested that permeability studies of rocks fractured at \textit{in situ} conditions through triaxial compression or direct shear tests can better represent fracture properties in the subsurface than prefractured or artificial specimens \citep{zhang2013laboratory,nygaard2006brittle,carey2015fracture,frash2016high,su2017experimental,klugesustainability}. Indeed, some factors such as the influence of  stress state during fracture, the transition from brittle to ductile failure, effective stress and hysteresis effects can be only investigated under \textit{in situ} conditions. An experimental campaign on shale fracture permeability using triaxial direct-shear coreflood methods provides a unique opportunity to continually measure permeability during the entire loading phase under \textit{in situ} conditions \citep{carey2015fracture,carey2015experimental,frash2016high,frash2017caprock}. 

Integrating experimental studies on shale fracture-permeability behavior into a numerical model is meaningful from both theoretical and practical points of view, considering that characterization of fracture permeability is difficult and costly in the field and in the laboratory. Modeling of fracture permeability remains a challenging problem in the following three aspects: (1) description of  cracking and fracturing events considering material heterogeneity; (2) proper selection of a fluid solver and an efficient coupling scheme to simulate coupled fluid flow in the damaged material; (3) capability of dealing with scaling effects. 

Continuum-based approaches, in which the material domain of interest is treated as a homogeneous continuum body or a multiphase continuum mixture if the materials are fluid-infiltrated, are widely used, %% in the literature \citep{}
but are inherently incapable of dealing with material heterogeneity and they fail to a large extent to simulate appropriately damage localization and the resulting size-effect. In these approaches, the fracture surfaces are usually represented by various methods, including but not limited to, embedded discontinuities \citep{mohammadnejad2013extended,salimzadeh2015three,jin2017fully}, phase field models \citep{mikelic2015phase,miehe2015minimization,choo2018coupled},
%Weixin: What do you mean with "status of the material"?
%Cusatis: In phase field model, they claimed that status of the material was used as a alternative to displacement jump to represent cracks
or smeared crack bands \citep{chau2016growth,li2017spherocylindrical}. 

On the contrary, approaches forfeiting the continuum assumption (also called discrete models), such as the discrete element method \citep{cundall1979discrete,potyondy2004bonded} and lattice/particle models \citep{cusatis2011lattice2,cusatis2011lattice,cusatis2003confinement,cusatis2003confinement2,smith2014discrete,li2017multiscale,ashari2017lattice,kozicki2008new,belheine2009numerical,bolander1998fracture,nikolic2016discrete}, considers the material domain as a three-dimensional system of interacting cells or particles often connected through lattice elements. These approaches come with a natural representation of fractures and cracks by discontinuities between adjacent cells or particles and, if formulated at the appropriate length scale, are able to reproduce critical features of material heterogeneity. 

Furthermore, the fluid transport and hydro-mechanical coupling effect can be simulated at different length scales. At the microscopic level, fluid flow can be solved by Navier-Stokes equations with certain assumptions and the hydro-mechanical coupling effect can be captured via force-based interaction across the fluid-solid interface \citep{catalano2014pore,robinson2014fluid}. Although conceptually simple, the microscopic approach requires a knowledge of the pore network structure and places a high demand on computational resources. An alternative to the costly microscopic approach is a mesoscopic/macroscopic approach \citep{bolander2004simulation,grassl20152d,wang2016semi,li2017discrete,li2017computational} which adopts diffusion equations to describe fluid flow in a porous medium and relies on phenomenological laws such as Darcy's law, Fick's law, and lubrication theory. By introducing diffusion coefficients, such as hydraulic conductivity, permeability, and diffusivity in the sense of homogenization over the multiphase mixture, this approach benefits from affordable computational cost. The choice of the length scale and the corresponding flow solver involves a trade-off between computational cost and physical realism. 

Finally, discrete models can reproduce well size effect phenomena. It has been widely demonstrated that various localization events (tensile strain softening, shear band, compaction band, etc.) result in a strong size effect, that is the dependence of macroscopic material properties (such as, for example, strength and fracture toughness) on sample size \citep{bavzant1984size,bazant1997fracture,li2017characterization,jin2017elastic2}. However, not enough attention has been paid to the effect of sample size on rock fracture permeability. Indeed, this aspect is extremely important for models that are calibrated and validated on the basis of laboratory measurements and then used for field applications. 

In this work, a three dimensional (3D) numerical model based on a Multiphysics Lattice Discrete Particle Model (M-LDPM) framework is formulated to study shale fracture permeability behavior by simulating triaxial direct-shear experiments.  
\section{Multiphysics Lattice Discrete Particle Model (M-LDPM) for shale}\label{sec:MLDPM}
M-LDPM is formulated in a discrete poromechanics setting by adopting two coupled dual lattices simulating mechanical and transport behaviors, respectively. The model adopts an ``a priori'' discretization of the internal structure of the material at the meso-scale, which is the length scale of major material heterogeneities.
%
%The geometric characterization of the material mesostructure and the generation of dual lattices is described in Section \ref{sec:duallattice}; the formulation of the mesoscale mechanical lattice model is presented in Section \ref{sec:mechanical}; and the derivation and implementation of diffusion equations for fluid flow is described in Section \ref{sec:fluid}.
%
%
\subsection{Dual lattice topology}\label{sec:duallattice}
\begin{figure*}
\centering
\includegraphics[width=0.85\textwidth]{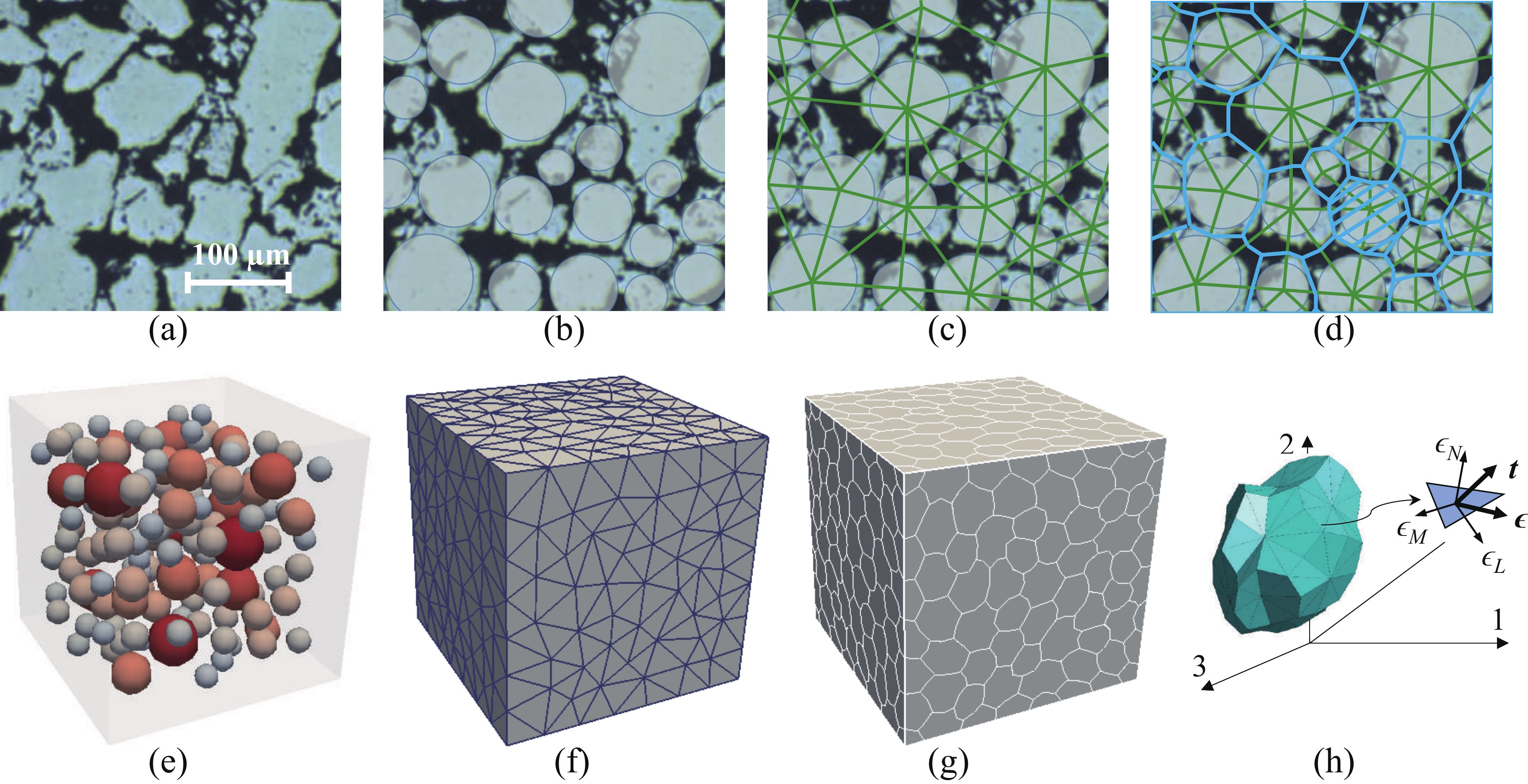}
\caption{(a) Granular internal structure of Toarcian shale \citep{akono2016microscopic}. Spherical particles placed at the center of shale grains in (b) 2D and (e) 3D domains. (c) 2D and (f) 3D representations of Delaunay tetrahedralization connecting spherical particles. (d) 2D and (g) 3D representations of a domain tessellation. (h) A 3D polyhedral cell representing a shale grain. }
\label{fig:LDPM-M}     
\end{figure*}

Typical rock materials feature an internal granular structures. For shale, grains consist of agglomerated clay particles and other minerals such quartz and feldspar, and in most cases are characterized by a maximum grain size in the order of \SI{50}{\um} as one can see in Fig. \ref{fig:LDPM-M}a with reference to the Toarcian shale \citep{akono2016microscopic}. 
%Weixin: include in Fig. \ref{fig:LDPM-M}a the length scale
%will do
In the M-LDPM formulation, the grain-scale heterogeneity of shale is replicated by discretizing the material domain with a system of randomly distributed polyhedral cells where a polyhedral cell represents a binder-coated grain. Only the coarser grains are considered in order to reduce computational cost while preserving the main source of heterogeneity. 

The granular lattice system is constructed through the following steps: (1) An artificial supporting system with spherical particles placed at the center of shale grains inside a certain volume of material is generated by a try-and-reject procedure devised to prevent mutual overlapping of particles and overlapping with the external boundary (see Fig. \ref{fig:LDPM-M}b for 2D representation and Fig. \ref{fig:LDPM-M}e for a 3D system). Details of the procedure can be found in \cite{cusatis2011lattice}. The size distribution of the spherical supports follows a fractal Cumulative Distribution Function (CDF): $P(d)=[1-(d_0/d)^q]/[1-(d_0/d_a)^q]$ where $d$, $d_0$, $d_a$ are particle size, minimum particle size, maximum particle size, respectively. In this work $q=2.5$ was used in all calculations. %The generated supporting particles are randomly distributed across the material domain on vertices, surface faces, and interior volume. 
Particle diameters $d_i$ are calculated by sampling the CDF. New particles are generated until the total volume of the generated spherical particles, $\tilde{V}_{a0} = \sum_i(\pi d_i^3/6)$, exceeds $V_{a0}$, where $V_{a0} = v_{a0}V$ is the total volume of the simulated particles, $v_{a0} = [1-(d_0/d_a)^{3-q}]v_a$ can be computed from the particle volume fraction $v_a$, and V denotes the specimen volume. Note that $d_0$, $d_a$, and $v_a$ need to be calibrated against the grain size distribution of the simulated granular rock.
%Weixin: here you need to describe how we determine the number of particles to be placed and write a few sentences about the placement procedure. See Shiva's paper.
%Cusatis: add a few sentences
The set of particles inside the interior volume is also augmented through surface nodes with zero radius and a given spacing to describe the external surface. 
(2) A constrained Delaunay tetrahedralization connects the given particle centers and fills the space without gaps. For 2D illustration, the tetrahedralization corresponds to a triangular mesh, as shown in Fig. \ref{fig:LDPM-M}c, whereas in an actual 3D material domain, a mesh of tetrahedra is created, as shown in Fig. \ref{fig:LDPM-M}f. (3) A 3D domain tessellation, anchored to the Delaunay tetrahedralization but not coinciding with the classical Voronoi tessellation, subdivides the domain into a system of polyhedral cells (see Fig. \ref{fig:LDPM-M}d for 2D representation and Fig. \ref{fig:LDPM-M}g for the 3D system). A generated polyhedral cell, as illustrated in Fig. \ref{fig:LDPM-M}h, represents a rock grain. The surface of each polyhedral cell is subdivided into triangles, called hereinafter ``facets'' and describing potential crack locations (see Fig. \ref{fig:LDPM-M}e). Each facet has one vertex inside the tetrahedron (tet-point), one on each edge of the tetrahedron (edge-point), and one on a face of the tetrahedron (face-point). The tessellation procedure is described in detail in \cite{cusatis2011lattice} and it is formulated in a way to minimize the intersection between each facet and the grains. This is done so that, as often verified in practice, cracks are simulated to occur at the grain interface and in the embedding matrix as opposed to cutting through the grains.

\begin{figure*}
\centering
\includegraphics[width=0.6\textwidth]{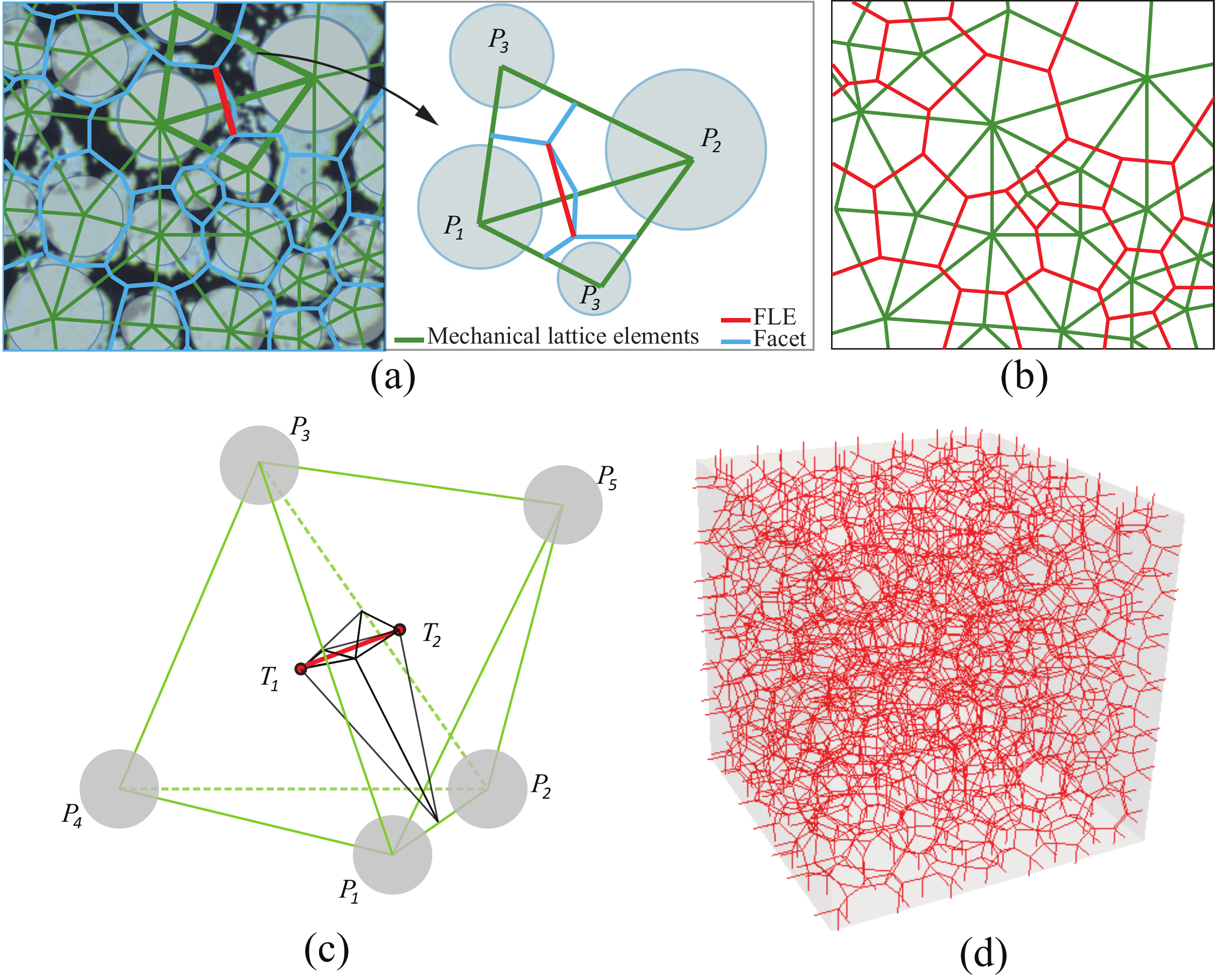}
\caption{ Generation of a Flow Lattice Element (FLE) connecting the tet-points belonging to two adjacent triangles/tetrahedra in (a) 2D and (c) 3D domain. (b) 2D and (d) 3D representation of a FLE network. }
\label{fig:FLE}     
\end{figure*}

While the LDPM cell system provides a geometrical characterization of the solid portion at the meso-scale and it serves as a structure for the description of the mechanical behavior, a dual lattice system allows for the coupled simulation of fluid flow at the same length scale. The dual lattice system is constructed as follows. Let's consider the 3D mesh of tetrahedra generated in step (2) as described above and its 2D representation as depicted in Fig. \ref{fig:FLE}a. A Flow Lattice Element (FLE) connects two points inside two adjacent tetrahedra (tet-points), which are also vertices of the corresponding polyhedral cells, as shown in Fig. \ref{fig:FLE}a for the 2D representation and Fig. \ref{fig:FLE}c for the 3D system. The fluid mass transport between these two tetrahedra is described by the FLE. Hence, the fluid flow across the entire computational domain can be represented by a network of FLE connecting all tetrahedra, as illustrated in Fig. \ref{fig:FLE}b and Fig. \ref{fig:FLE}d. In addition, a thin layer of elements orthogonal to the external surface of the domain is generated to facilitate application of the boundary conditions needed for solving flow problems. 

% Weixin: In figure (c) I would only visualize the 6 facets that are associated  with the shown FLE. Otherwise it is very difficult to understand. 
%will do

It is worth pointing out that the term ``dual lattice'' represents the topological relation between the LDPM cell system for the solid phase and the network of FLE for the fluid phase. This is similar to the dual-graph concept adopted by \cite{grassl20152d,grassl2016three,ulven2017capturing}. Under such an arrangement, the flow/transport elements are aligned in the general direction of potential crack paths, and thus the influence of local cracks on the fluid flow, and vice versa, the influence of the fluid pressure on the solid, can be captured naturally. With reference to Fig. \ref{fig:FLE}a (for 2D) and \ref{fig:FLE}c (for 3D), one can see that each FLE is associated to a certain characteristic number of facets (2 in 2D and 6 in 3D) belonging to the two LDPM cells associated to the mechanical lattice elements $P_1-P_2$ for 2D, and $P_1-P_2$, $P_1-P_3$, $P_2-P_3$ for 3D.
%
%belonging to two particles $P_1$ and $P_2$ are the potential crack path. The projection of the interfaces into a plane orthogonal to the triangle edge (mechanical strut) connecting $P_1$ and $P_2$ ends up with the projected interfaces aligning with the corresponding FLE. Similarly, in the 3D domain, each FLE can be associated with six facets of the LDPM cells. 
%
This makes possible the seamless two-way coupling of the mechanical and flow behaviors as will be presented later in this paper.
%Note that different from the models by \cite{grassl20152d,grassl2016three}, the mechanical and fluid flow model in this work are both formulated by using a discrete setting, which will be discussed in the following sections. 

\subsection{LDPM for shale mechanical behavior}\label{sec:mechanical}
\subsubsection{Kinematics}
In the LDPM formulation, adjacent cells interact through shared triangular facets. Rigid-body kinematics is adopted to describe the heterogeneous deformation of the cell system. Three strain measures, one normal component and two shear components, are defined at each facet (see Fig. \ref{fig:LDPM-M}h) as
\begin{equation}\label{eq:strain}
\epsilon_N = \frac{\mathbf{n}^T \llbracket {\mathbf{u}_{C}} \rrbracket}{\ell}; \quad \epsilon_M = \frac{\mathbf{m}^T \llbracket {\mathbf{u}_{C}} \rrbracket}{\ell}; \quad \epsilon_L = \frac{\mathbf{l}^T \llbracket {\mathbf{u}_{C}} \rrbracket}{\ell}
\end{equation}
where $\llbracket {\mathbf{u}_{C}} \rrbracket$ is the displacement jump vector calculated by the displacements and rotations of the nodes adjacent to the selected facet; $\ell$ = tetrahedron edge associated with the facet; $\mathbf{n}$ is a unit vector normal to each facet, and $\mathbf{m}$ and $\mathbf{l}$ are two mutually orthogonal unit vectors contained in a plane orthogonal to $\mathbf{n}$. It is worth pointing out that (1) to avoid unrealistic shear locking phenomena \citep{cusatis2011lattice}, the local system of reference is defined on the projection of the LDPM facets in a plane orthogonal to the lattice connecting the two cells; and (2) the strain definitions in Eq. \ref{eq:strain} correspond to the projection into the facet system of reference of the classical strain tensor \citep{cusatis2017discontinuous,cusatis2013high,rezakhani2016asymptotic}.

\subsubsection{LDPM constitutive laws}
The facet stress vector applied to the solid phase,  $\mathbf{t}_s = t_N \mathbf{n} + t_M \mathbf{m} + t_L \mathbf{l}$, is calculated through appropriate constitutive laws, which describes various phenomena of the solid grain interaction (Fig. \ref{fig:LDPM-M}g). 

In the elastic regime, the mechanical facet stress components are proportional to the corresponding strain components:
\begin{equation}
{t}_\beta = E_\beta\epsilon_\beta~~~(\beta=N,M,L)
\end{equation}
%${t}_N = E_N\epsilon_N$; ${t}_M = E_T\epsilon_M$; ${t}_L = E_T\epsilon_L$,
where $E_N$ is the normal modulus, $E_M =E_L=\alpha E_N$ is the shear modulus, and $\alpha =$ shear-normal coupling parameter. 
%
%$E_N$ and $\alpha$ can be related to macroscopic elastic properties for isotropic materials according to the following relation
%\begin{equation}\label{eq:micromacro}
%E_N = \frac{E}{1-2\nu}; \qquad \alpha = \frac{1 - 4\nu}{1+\nu}
%\end{equation}
%where $E$ and $\nu$ are Young's modulus and Poisson's ratio of the solid phase, respectively. 

Beyond the elastic limit, the vectorial constitutive laws are formulated to reproduce two distinct nonlinear phenomena.

The first source of nonlinearity is related to fracturing and cohesive behavior under tension and tension/shear that occur for $\epsilon_{N}>0$. Following the work of \cite{cusatis2003confinement,cusatis2003confinement2},
one can define an effective strain and an effective stress as $\epsilon =[\epsilon_{N}^{2}+\alpha (\epsilon_{M}^{2} + \epsilon_{L}^{2})]^{1/2}$, $t=[{t}_{N}^2+({t}_{M}^2+{t}_{L}^2)/\alpha]^{1/2}$, respectively, and compute the normal and shear stresses as follows:
\begin{equation}
{t}_{N}=\epsilon_{N}(t/\epsilon); ~~{t}_{M}=\alpha \epsilon_{M}(t/\epsilon); ~~{t}_{L}=\alpha \epsilon_{L}(t/\epsilon)
\end{equation}
The effective stress $t$ is incrementally elastic ($\dot{t}=E_N \dot{\epsilon}$) and must satisfy the inequality $0 \leq t \leq \sigma_{bt}(\epsilon,\omega)$, in which 
\begin{equation}
\sigma_{bt} = \sigma_0(\omega) \exp \left[-H_0(\omega) \langle \epsilon_{\max}- \epsilon_0(\omega)\rangle/\sigma_0(\omega) \right]
\end{equation}
%$\sigma_{bt} = \sigma_0(\omega) \exp \left[-H_0(\omega) \langle \epsilon_{\max}- \epsilon_0(\omega)\rangle/\sigma_0(\omega) \right]$, 
where $\langle x\rangle=\max(x,0)$, $\tan(\omega)=\epsilon_{N}/(\sqrt{\alpha} \epsilon_{T}$, and $\epsilon_T = [\epsilon_M^2 + \epsilon_L^2]^{1/2}$. The function $\sigma_0(\omega)$ represents the strength limit for the effective stress, and is given by 
\begin{equation}
\sigma_{0}(\omega) =\sigma_{t} \frac{-\sin(\omega)+ \sqrt{\sin^2(\omega)+4 \alpha \cos^2(\omega)  r_{st}^2}}{2 \alpha \cos^2(\omega) r_{st}^2}
\end{equation}
where $r_{st} = \sigma_s/\sigma_t$ is the ratio between the mesoscale shear strength (or cohesion) $\sigma_s$ and the mesoscale tensile strength $\sigma_t$. After the maximum effective strain reaches its elastic limit $\epsilon_0(\omega) = \sigma_0/E_N$, the stress boundary $\sigma_{bt}$ decays exponentially with a softening modulus, $H_0 (\omega) = H_t (2\omega/\pi)^{n_t}$, which provides a smooth transition from softening behavior under pure tension ($\omega = \pi/2$), $H_0 = H_t$, to perfectly plastic behavior under pure shear ($\omega = 0$), $H_0 = 0$. In most cases, $n_t = 0.2$. For the correct energy dissipation during mesoscale damage localization to be preserved, the softening modulus in pure tension is expressed as $H_t = 2 E_N /(\ell_t /\ell -1)$, where $\ell_t = 2E_N G_t / \sigma_t^2$ is the characteristic length, $G_t$ is the mesoscale fracture energy, and $\ell$ is the length of the tetrahedron edge (lattice element) associated with the facet. 

Since relevant to the the effect of cracking on permeability discussed later, it is worth pointing out that the mesoscale crack opening components (normal and shear) can be calculated as 
%$\mathbf{w} = w_N \mathbf{n} + w_M \mathbf{m} + w_L \mathbf{l}$, where
\begin{equation}
\delta_\beta = \ell(\epsilon_\beta - {t}_\beta/E_\beta)~~~(\beta=N,M,L)
\label{eq:wn}
\end{equation}

%\begin{subequations}
%\begin{align}
%w_N& = \ell(\epsilon_N - {t}_N/E_N);\label{eq:wn}\\
%w_M& = \ell(\epsilon_M - {t}_M/E_T);\\
%w_L& = \ell(\epsilon_L - {t}_L/E_T)
%\end{align}
%\end{subequations}
%in which $w_N$ is its normal component with positive and negative values representing opening and closure of cracks, respectively, and $w_M$ and $w_L$ are two shear components related to sliding between crack surfaces.  

The second source of nonlinearity is related to frictional behavior under compression/shear for $\epsilon_N < 0$. This can be simulated effectively through a non associative incremental plasticity formulation in which, during plastic flow, $\varphi(t_N, t_M, t_L)=0$, the incremental shear stresses are computed as $\dot{{t}}_M = E_T (\dot{\epsilon}_M - \dot{\epsilon}_M^P)$, $\dot{{t}}_L = E_T (\dot{\epsilon}_L - \dot{\epsilon}_L^P)$ and the normal stress is simply elastic $\dot{{t}}_N = E_N \dot{\epsilon}_N$. In the previous expressions, $\varphi(t_N, t_M, t_L)$ is the yielding function,  $\dot{\epsilon}_M^P = \dot{\lambda} \partial \psi /\partial {t}_M$ and $\dot{\epsilon}_L^P = \dot{\lambda} \partial \psi /\partial {t}_L$ are the shear plastic strain increments, and $\lambda$ is the plastic multiplier, $\psi=\psi_0[{t}_M^2 + {t}_L^2]^{1/2}$ is the plastic potential. The yielding function can be formulated according to a Mohr-Coulomb criterion as 
\begin{equation}
\varphi = [{t}_M^2 + {t}_L^2]^{1/2} - \sigma_s + \mu_0 t_N
\end{equation}
in which $\mu_0$ is the internal friction coefficient.  

%In addition, an elastic behavior is assumed for the normal response, that is, ${t}_N = E_N \epsilon_N$ for $\epsilon_N < 0$. 
%\textcolor{blue}{Note from Bill: It might be helpful to have a plot showing the stress-displacement behavior as described in the above paragraphs.}
%
%The LDPM elastic behavior is statistically isotropic if $E_N$ and $\alpha$ are constant. Various degrees of anisotropy can be also included by assuming $E_N$ and $\alpha$ to vary with the facet orientation \citep{}.
% Include here your shale paper and the microplane paper with Congrui. Also Bazant and Brocca paper.

The facet stresses calculated through the constitutive laws described above represent the stresses carried by the solid phase. Equilibrium considerations at the facet level allow for the reasonable assumption of a parallel coupling between the stresses carried by the solid phase and those by the fluid phase. In this work, the effective stress concept is adopted, and the total stress vector on each facet can be computed as 
\begin{equation}\label{eq:effsig}
\mathbf{t} = \mathbf{t}_s - b \mathbf{t}_w
\end{equation}
where $b$ is the Biot coefficient, and $\mathbf{t}_w = p \mathbf{n}$ is the fluid pressure vector. The negative sign in Eq. \ref{eq:effsig} comes from the pressure sign convention, which is positive for the fluid and negative for the solid.

At this point a few comments are in order.
\begin{enumerate}
\item The constitutive equations described above depend on 6 parameters that must be identified by fitting macroscopic experimental data relevant to elastic behavior (for $E_N$ and $\alpha$), tensile fracturing (for $\sigma_t$ and $\ell_t$), unconfined compression and triaxial compression (for $\sigma_s$ and $\mu_0$).
\item The resulting macroscopic mechanical response is statistically isotropic. Material anisotropy, typical of many shale formations, can be accounted for by postulating the dependence of the above parameters on the facet orientation \citep{jin2017elastic} and by explicitly simulating bedding planes. This was already done by \cite{li2017multiscale} and it is not adopted here simply because the particular shale samples analyzed in this paper do not show appreciable anisotropy.
\item The constitutive equations do not include nonlinearity for high confining pressures under hydrostatic loading, pressure dependence on the elastic behavior, and creep behavior. These aspects, albeit important, are outside the scope of this study.
\item Equation \ref{eq:effsig} is formulated based Biot's theory of poroelasticity \citep{biot1941general} and for intact shale one can assume $b\approx0.6$ \citep{suarez2013evaluating}. %include relevant citation
Beyond the elastic regime, the effective stress concept still holds, but the Biot coefficient needs to evolve with the level of damage \citep{chau2016growth} and must be equal to one ($b=1$) at the location of a stress free crack ($\mathbf{t}_s=\mathbf{0}$). Such formulation, however, requires calibration with data not available in the current study. Hence the constant value $b=0.6$ will be used in all simulations. It must be mentioned that this assumption does not reduce the validity of the results because in the analyzed experiments the fluid pressure is much smaller than the stresses on the solid phase ($|\mathbf{t}_s| \gg p$), making the dependence of the response on $b$ negligible.
\end{enumerate}

\subsubsection{Equilibrium}
%
%Although researchers were trying to develop hypothetical formulas to relate the Biot coefficient with damage variables , such formulas need to be calibrated against experimental data requiring measurement of the Biot coefficient as a function of fracturing processes, which, however, are not generally available. The formulation of the Biot coefficient is beyond our current scope, and a proper value will be selected depending as required. 
%
The equilibrium equations are obtained through the force and moment equilibrium of each cell subject to the force resultants obtained by multiplying the tractions in Eq. \ref{eq:effsig} times the facet areas for all facets belonging to the given cell. In absence of body forces, one can write 
\begin{equation} \label{motion-1}
M_u^I\ddot{\mathbf{U}}^I =\sum_{\mathcal{F}_I} A_{f} \mathbf{t}^{IJ}
%M_u^I\ddot{\mathbf{U}}^I - V^I \mathbf{b}^0=\sum_{\mathcal{F}_I} A_{\mathcal{F}} \mathbf{t}^{IJ}
~~~M_{\theta}^I\ddot{\mathbf{\Theta} }^I = \sum_{\mathcal{F}_I}  A_{f}  \mathbf c^I \times \mathbf {t}^{IJ}
\end{equation}
%and
%\begin{equation}\label{motion-2}
%M_{\theta}^I\ddot{\mathbf{\Theta} }^I = \sum_{\mathcal{F}_I}  A_{\mathcal{F}}  \mathbf c^I \times \mathbf {t}^{IJ}
%\end{equation}
where $\mathcal{F}_I$ is the set of facets surrounding the node $P_I$ (located inside cell $I$) and obtained by collecting all the facets associated with each adjacent cell pairs $(I,J)$; $A_{f}$ = projected facet area; %$V^I$ is the cell volume; 
$\mathbf{c}^I$ = vector connecting the node $P_I$ to the facet centroid; %$\mathbf{b}^0$ is the body force vector; 
$M_u^I$, $M_\theta^I$ = mass and moment of inertia of cell $I$; and $\ddot{\mathbf{U}}^I$, $\ddot{\mathbf{\Theta} }^I$ are acceleration and rotational acceleration, respectively, of cell $I$. In the current implementation, an explicit dynamic algorithm (based on a central difference scheme) is adopted to solve the equations above \citep{cusatis2011lattice}. This offers the advantage of avoiding the convergence problems that implicit schemes often have in handling softening behaviors. 

\subsection{Discrete formulation of fluid flow}\label{sec:fluid}
This section discusses the formulation of flow phenomena in a discrete setting consistent with the LDPM framework for shale introduced earlier. Attention is restricted to the flow of water under full saturation conditions, constant room temperature, and assuming that water behaves as a slightly compressible Newtonian fluid.  

\begin{figure}
\centering
\includegraphics[width=0.45\textwidth]{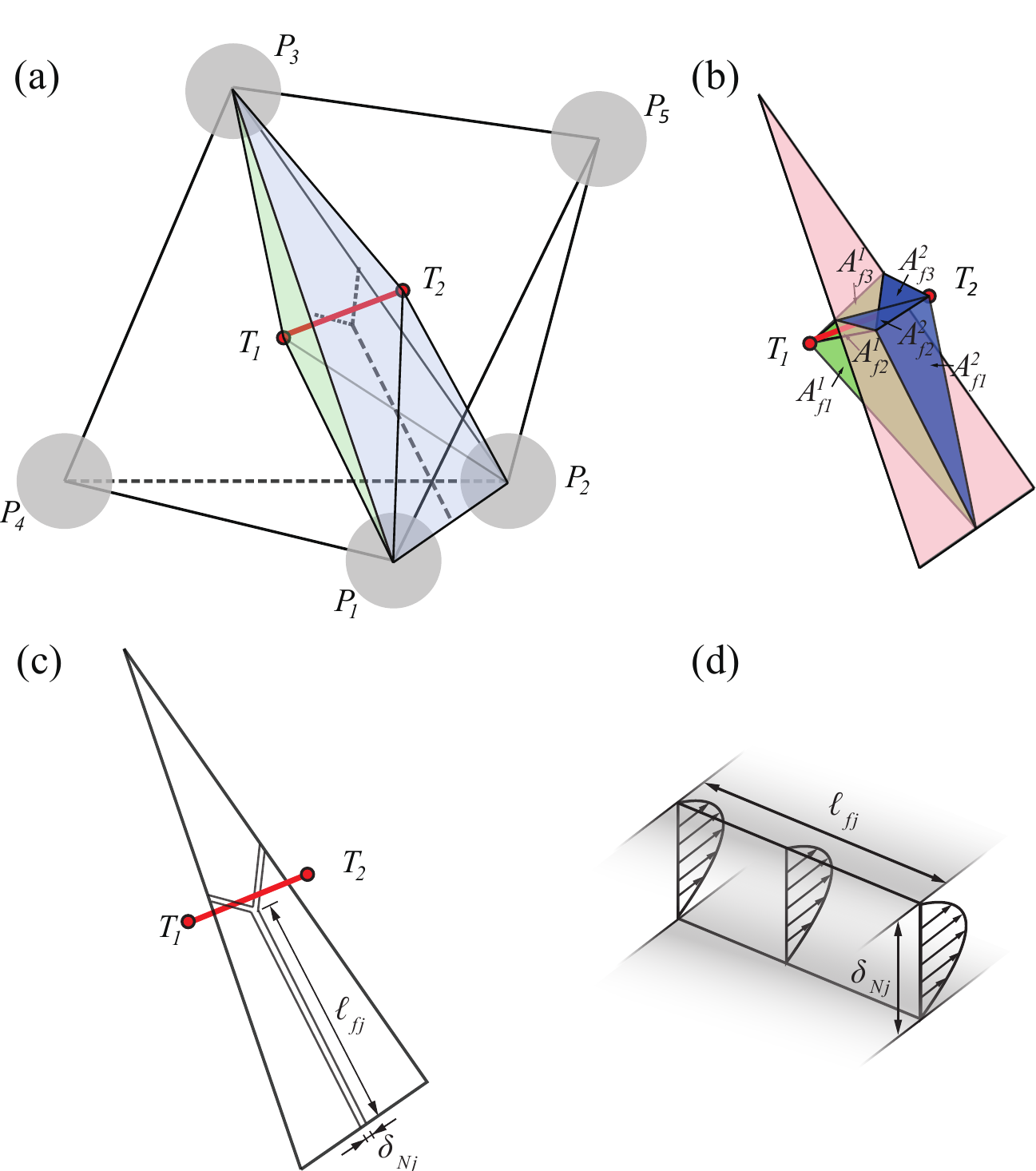}
\caption{(a) A Flow Lattice Element (FLE) and its associated control volumes, (b) the associated LDPM facets, (c) the cracked triangle face and the illustration of normal crack opening, (d) Poiseuille flow along the cracked surfaces. }
\label{fig:Flow}     
\end{figure}

Let's consider two adjacent tetrahedra, each connecting four LDPM cells in the undeformed configuration: $P_1$, $P_2$, $P_3$, $P_4$ and $P_1$, $P_2$, $P_3$, $P_5$ in Fig. \ref{fig:Flow}a. They have a common triangular face $\triangle_{123}$ across which a FLE connects the tet-points $T_1$ and $T_2$ located inside the two tetrahedra. The FLE is labeled $E_{12}$ and it is associated to two pyramidal volumes, $V_1$ and $V_2$, identified by the points $P_1$, $P_2$, $P_3$, $T_1$ and $P_1$, $P_2$, $P_3$, $T_2$.  The volumes can be computed as $V_1 = A_nL_1/3$, and $V_2 = A_nL_2/3$, where $A_n=A\mathbf{a}^T \mathbf{e}$ is the area calculated by projecting the original area of the face $\triangle_{123}$, $A$, into a plane orthogonal to $E_{12}$; $\mathbf{a}$ is the unit vector orthogonal to the face $\triangle_{123}$; $\mathbf{e}$ is the unit vector oriented as $E_{12}$; $L_1$, $L_2$ are the distance between point $F$ and $T_1$, $T_2$, respectively; point $F$ is the point where the triangular face $\triangle_{123}$ intersects $E_{12}$. It is worth noting that $L_1 + L_2 = L$, where $L$ is the length of $E_{12}$. If one defines the fraction coefficients $g_1=L_1/L$ and $g_2=L_2/L$ ($g_1 + g_2 = 1$), one obtains $V_1 = g_1 V$, and $V_2 = g_2 V$, in which $V = A_n L/3 = V_1+V_2$ represents the total volume of material associated with $E_{12}$. %Mass balance is considered for $V_1$ and $V_2$ with the interface $\triangle_{123}$. 
%
%Similar to the mechanical formulation as discussed in Section \ref{sec:mechanical}, a discrete setting is adopted in the formulation of the governing equations for fluid flow, with which discrete variable is assigned to the flow network node located at the tet-node of each tetrahedron. 

\subsubsection{Water content and flux for uncracked shale}
%
%Consider the volume $V_1$ to be a control volume. 
With reference to the control volume $V_i$ ($i$=1,2), the mass of water can be written as $M_{wu}^{i} =  m_{wi} V_i$ in which $m_{wi}$ is the water mass content, defined to be the water mass per unit reference volume (see, among others, \cite{rice1976some,wang2000theory}). The change in water mass content can be related to the increment of water content $\zeta^i = (m_{wi} - m_{w0})/\rho_{w0}$, where $m_{w0}$ and $\rho_{w0}$ are water mass content and density in the reference state, respectively. For slightly compressible fluids, the water density in the current state can be related to $\rho_{w0}$ by defining the bulk modulus $K_w$; one can write
\begin{equation}
\rho_{wi}=\rho_{w0}\left( 1+\frac{p_i-p_0}{K_w} \right)~~~(i=1,2)
\end{equation}
where $p_i$ is the fluid pressure in $V_i$, and $p_0$ is the initial/reference pressure.
%Weixin: This formula gives the reference for zero pressure ... shouldn't it actually be the initial/reference pressure p_{i0}?
%Cusatis: You're right. I changed it. 

According to the classic theory of poromechanics \citep{biot1941general,rice1976some,detournay1993fundamentals}, the increment of water content, $\zeta_i$, can be expressed as a linear combination of the volumetric strain, $\varepsilon_i$, of the solid phase
% Weixin: what is the medium? solid? homogenized solid+fluid? 
% Cusatis: volumetric strain was defined on the solid skeleton (include pore volume)
 defined as the relative variation of the solid volume, and the water pressure, $p_i$, as $\zeta_i = b\varepsilon_{i} + p_i/M_b$, where $M_b$ denotes the Biot modulus (also defined as the reciprocal of the so-called storage coefficient).
% Weixin: is M_b constant? should it also vary with damage? Shouldn't we comment on this
% Cusatis: In theory, M_b should also vary with damage. We can comment on this in the next subsection. 
It is worth pointing out that $b$ and $M_b$ may vary due to material hetereogeneity. The effect of this variation is insignificant in the context of this paper and will be neglected thereinafter. Also, the previous discussion on $b$ and its evolution with damage also holds for $M_b$. Again, in absence of relevant experimental data a constant value is actually used in this paper.

%Given the fluid mass in the uncracked control volume, $M_{fe}^{(1)}$, and the fluid mass flux into the same volume through the face$\triangle_{123}$, $Q_e^{(1)}$, the mass balance can be written as $\dot{M}_{fe}^{(1)} = Q_e^{(1)}$ in the absence of internal source or sink, where the superimposed dot represents time derivative, and the superscript $^{(1)}$ represents the volume $V_1$. With a poromechanics approach, the fluid mass $M_{fe}^{(1)}$ can be express as $M_{fe}^{(1)} =  m_f V_1$, in which $m_f$ is fluid mass content, defined to be the fluid mass per unit reference volume (see \cite{rice1976some,wang2000theory}). The change in fluid mass content can be related to increment of fluid content $\xi$ by $\xi = (m_f - m_{f_0})/\rho_{f_0}$ in the poroelastic regime, where $m_{f_0}$ and $\rho_{f_0}$ are the fluid mass content and the fluid density in the reference state, respectively. The fluid density in the current state can be related to $\rho_{f_0}$ by defining the bulk modulus of the fluid as $K_f = \rho_{f_0} p/(\rho_f - \rho_{f_0})$. According to the classic theory of poroelasticity \citep{biot1941general,rice1976some,detournay1993fundamentals}, the increment of fluid content $\xi$ can be expressed as a linear combination of the volumetric strain of the medium, $\varepsilon_{v}$ defined as the relative variation of the volume, and the fluid pressure, $p$, i.e. $\xi = b\varepsilon_{v} + 1/M_b p$, where $M_b$ denotes Biot modulus, and is the reciprocal of constant strain storage coefficient. 
%
One can write the time variation of the water mass in the control volume as
\begin{equation}\label{eq:fluidmass}
\dot{M}_{wu}^{i} =  \rho_{w0} \left( b \dot{\varepsilon}_{i} + \frac{\dot{p_i}}{M_b}\right) V_i~~~(i=1,2)
\end{equation}
%

%Similarly, $\dot{M}_2$ for the volume $V_2$ reads
%\begin{equation}\label{eq:fluidmass}
%\dot{M}_2 =  \rho_{f_0} \left( b\frac{\partial \varepsilon_{kk}}{\partial t} + \frac{1}{M}\frac{\partial p}{\partial t}\right) V_2
%\end{equation}
The mass flux through the face $\triangle_{123}$ from $V_1$ into $V_2$ can be obtained by using Darcy's law, which can be written as
\begin{equation}\label{eq:Q1}
Q_{wu}^{12} = \bar{\rho}_{w} \frac{\kappa_0}{\mu_w}A_n\frac{p_1-p_2}{L}
\end{equation}
where $\kappa_0$ and $\mu_w$ denote the intrinsic permeability of the material and the water viscosity, respectively; $p_1$, $p_2$ are the values of fluid pressure at points $T_1$ and $T_2$, respectively; and $\bar{\rho}_w=g_1 \rho_{w1}+g_2 \rho_{w2}$ is an estimate of the average density in the volume $V$.
%Weixin: is this what you used or the average?
%Cusatis: we used the average value eventually 
%Weixin: These formulas are more correct and since I do not think the difference on the results is big, let's keep it like it is now.
%
%Note that the pressure gradient in the Darcy's law is approximated by a finite difference scheme thanks to the discrete setting of the model. Similarly, the flux into the uncracked volume 2 can be written as
%\begin{equation}\label{eq:Q2}
%Q_e^{(2)} = \rho_{f} \frac{\kappa_0^{(2)}}{\mu_f}A_n\frac{p_F-p_2}{L_2}
%\end{equation}
%where $\kappa_0^{(2)}$ denotes the intrinsic permeability of the material associated with $V_2$, and $p_2$ is pressure variable at node $T_2 $. 

%COMMENT HERE ON LIMITATIONS
Although Darcy's law is classically adopted to calculate flow rate through a porous material, as pointed out by \cite{falk2015subcontinuum,obliger2018impact}, it might not be completely appropriate to predict transport in shales due to the strong adsorption in kerogen and the breakdown of hydrodynamics at the nanoscale. Analysis of non-darcy behavior of shale is beyond the scope of the current paper. Moreover, this aspect does not significantly affect the results since shale fracture-permeability is mainly governed by the flow along cracked surfaces, as will be discussed below. 
% Cusatis: I'm not quite sure what you would like to comment on limitations
% Weixin: A few sentences related to the fact that Darcy's law might not be completely appropriate due to the hydrophobic character of the kerogen + citation of Ulm paper.
% Cusatis: added 

\subsubsection{Water content and flux for cracked shale}
The influence of local cracking can be addressed by introducing the contribution of the water mass stored in the cracked domain and the water flux along the crack surfaces. Let's consider again $E_{12}$ and the six associated LDPM facets of areas $A^i_{f1}$, $A^i_{f2}$, and $A^i_{f3}$ belonging to $V_i$ ($i$=1,2) as it is illustrated in Fig. \ref{fig:Flow}b. The water mass stored in the cracks is $M_{wc}^{i} = \rho_{wi} V_{ci}$, where the cracked volume can be expressed as $V_{ci} = \sum_{j=1}^3 A^i_{fj}\delta^i_{Nj}$ and $\delta^i_{Nj}$ are the normal crack openings. 

The time variation of the water mass in the cracks can be written as
\begin{equation}
\dot{M}_{wc}^{i}= \rho_{w0}\frac{V_{ci}}{K_w}\dot{p}_i  + \rho_{wi} \dot{V}_{c}^{i}
\end{equation}

%Note that the cracking opening $w_{Ni}$ of the $i^{th}$ facet can be obtained from the mechanical model according to Eq. \ref{eq:wn}. The total fluid mass stored in the volume $V_1$ is the sum of the one for the uncracked domain and the one for  the cracked domain, i.e. $M_{f}^{(1)} = M_{fe}^{(1)}+M_{fc}^{(1)}$. Correspondingly, its time derivative can be written as
%\begin{equation}
%\begin{split}
%\dot{M}_{f}^{(1)} &=  \rho_{f_0} \left( b\frac{\partial \varepsilon_{v}^{(1)}}{\partial t} + \frac{1}{M_b}\frac{\partial p_1}{\partial t}\right) V_1 \\
%&+ \frac{\rho_{f_0}}{K_f}\frac{\partial {p_1}}{\partial t} V_{c}^{(1)} + \rho_f \frac{\partial{V}_{c}^{(1)}}{\partial t}
%\end{split}
%\end{equation}

The water mass flux, $Q_{wc}^{12}$, from $V_1$ into $V_2$ associated with the cracks, can be approximated by assuming a steady laminar flow between two crack surfaces with a cross section of length $\ell_{fj}$ and width $\delta^i_{Nj}$ with $j = 1,...,3$, where $\ell_{fj}$ represents the intersection of the $j^{th}$ facet with the tetrahedron face $\triangle_{123}$ (see Fig. \ref{fig:Flow}c and d). In this case, the solution of a two dimensional Poiseuille flow in a channel \citep{massey1998mechanics}, known as Poiseuille's formula, can be adopted. 

%Note that only the flux component orthogonal to the tetrahedron face $\triangle_{123}$ is considered. The same assumption is also made for the flux into the volume $V_2$ associated with cracking, $Q_{c}^{(2)}$. As a consequence, $Q_{c}^{(1)}$ and $Q_{c}^{(2)}$ can be expressed as 
One can write
\begin{equation}
Q_{wc}^{12} = \bar{\rho}_{w} \frac{\kappa_c}{\mu_w}A_n\frac{p_1-p_2}{L}
\end{equation}
where 
\begin{equation}
\label{eq:kc}
\kappa_c=\frac{1}{12A}\left( \frac{g_1}{I_{c1}}+ \frac{g_2}{I_{c2}} \right)^{-1}
\end{equation}
%
%\begin{subequations}
%\begin{equation}\label{eq:Qc1}
%Q_{c}^{(1)} = \frac{\rho_f}{12 \mu_f} \mathbf{n}^T\mathbf{f} I_{c}^{(1)} \frac{p_F - p_1}{L_1}
%\end{equation}
%\begin{equation}\label{eq:Qc2}
%Q_{c}^{(2)} = \frac{\rho_f}{12 \mu_f} \mathbf{n}^T\mathbf{f} I_{c}^{(2)} \frac{p_F - p_2}{L_2}
%\end{equation}
%\end{subequations}
%
and $I_{ci} = \sum_{j = 1}^3 {\ell_{fj}}{(\delta^i_{Nj})^3}$ ($i$=1,2). Equation \ref{eq:kc} is derived assuming that the cracked permeabilities in $V_{c1}$ and $V_{c2}$ are coupled in series.

\subsubsection{Mass balance equations}
The total water mass and total water flux can be obtained by adding the contributions from the uncracked and cracked domains. In this case, the mass balance equations for volume $V_1$ and $V_2$ can be written as
\begin{equation}\label{eq:mass-bal-1}
\left(b \dot{\varepsilon}_{1} + \frac{\dot{p_1}}{M_b} \right)V_1 +\frac{V_{c1}\dot{p}_1}{K_w }  + \frac{\rho_{w1}\dot{V}_{c1}}{\rho_{w0}}  =  \xi \frac{A_n}{ L} (p_2-p_1)
\end{equation}
and
\begin{equation}\label{eq:mass-bal-2}
\left(b \dot{\varepsilon}_{2} + \frac{\dot{p_2}}{M_b} \right)V_2 +\frac{V_{c2}\dot{p}_2}{K_w }  + \frac{\rho_{w2}\dot{V}_{c2}}{\rho_{w0}}  =  \xi \frac{A_n}{ L} (p_1-p_2)
\end{equation}
where both equations are normalized with the reference water density $\rho_{w0}$, $\xi= \bar{\rho}_{w} (\kappa_0+\kappa_c)/(\rho_{w0}\mu_w)$, and the superimposed dot represents time derivative. 
%
%
%\subsubsection{Numerical implementation}
Equations \ref{eq:mass-bal-1} and \ref{eq:mass-bal-2} can be rewritten in matrix form as 
%
%The semi-discrete matrix form of Eqs. \ref{eq:go1} and \ref{eq:go2} for the element $E_{12}$ can be obtained by defining a vector of nodal variables $\mathbf{u}_e = \{ p_1, p_2 \}^T$, which reads
\begin{equation}
\mathbf{C}\mathbf{\dot{p}} + \mathbf{D} \mathbf{p} + \mathbf{S} = \mathbf{0}
\end{equation}
where $\mathbf{C}$ and $\mathbf{D}$ are the capacity (or storage) matrix and the conductance matrix of the FLE $E_{12}$, respectively; $\mathbf{S}$ is the source term; and $\mathbf{p} = [ p_1, p_2 ]^T$. One has
%These matrices have the following forms: 
\begin{equation}
\mathbf{C} = V
	\begin{bmatrix}
	g_1 c_{1} & 0\\
    0 & g_2 c_{2} 
	\end{bmatrix}
\end{equation}
\begin{equation}
    \mathbf{D} = \xi \frac{A_n}{L}
	\begin{bmatrix}
	 1& -1\\
    -1 & 1 
	\end{bmatrix}
\end{equation} 
\begin{equation}
    \mathbf{S} = V
	\begin{bmatrix}
	g_1 s_{1}\\
    g_2 s_{2}
	\end{bmatrix}
\end{equation}
where $c_i=M_b^{-1}+V_{ci}(K_wV_i)^{-1}$ and \\ $s_i=b\dot{\varepsilon_i}+\rho_{wi}\dot{V}_{ci}(\rho_{w0}V_i)^{-1}$.
%$C_f = \rho_{f_0}(1/M_b + 2/K_f (V_c^{(1)}/V_1 + V_c^{(2)}/V_2))$, $D_f = \rho_f \kappa_{eq}/\mu_f$, and 
%$S_f = -\rho_{f_0} b/2(\partial \varepsilon_{v}^{(1)}/\partial t + \partial \varepsilon_{v}^{(2)}/\partial t) + \rho_f/2 (\dot{V}_c^{(1)}/V_1 + \dot{V}_c^{(2)}/V_2)$. 
%
%
It is worth observing that, to reduce the memory requirements of the calculations, it is possible, without significant difference in the results, to substitute $c_i$ and $s_i$ ($i=1,2$) with the volume averages $\bar{c}=g_1c_1+g_2c_2$ and $\bar{s}=g_1s_1+g_2s_2$.
%
%Weixin: please check all equations above. I reorganized things. The balance equations are divided by the water reference density.
%Cusatis: I already checked. Corrected some typos.
%
%Note that the capacity term and the source term are averaged over $V_1$ and $V_2$ in order to simplify the implementation.
The global matrices of the overall governing equations of the flow problem are obtained by assembling the matrix contributions of all flow elements. 
It is worth mentioning that the contributions of the flow elements associated with the boundary layers are also included. Dirichlet type boundary conditions can be imposed on the external nodes of the boundary layers through the penalty method. The fluxes of the boundary layer elements can be also calculated. Thanks to their perpendicularity to the external surfaces, the sum of the fluxes
%each flux multiplied by the corresponding surface area 
(Riemann sum) provides an estimate of the volumetric flow rate over the boundaries. 
In this work, the time integration of the flow problem was performed by means of the Crank-Nicolson method \citep{di2009hygro,bavzant2017creep}.

% Weixin: discuss here the boundary layers and the BCs
% Cusatis: I added a few sentence for the BCs

Finally, the numerical implementation is completed by the application of a staggered coupling scheme with the solver of the mechanical problem. At the selected simulation time step, the mechanical model makes use of the solution (fluid pressure) from the fluid flow model, whereas the flow problem properties are updated by using the information from the mechanical model (crack opening and volumetric strain). The staggered coupling scheme along with the mixed explicit-implicit integration scheme provides a simple yet efficient approach for two-way coupling simulations of mechanical and flow behaviors.

The proposed framework is implemented into the MARS software \citep{pelessoneMARS}, which is a structural analysis computer code with an object-oriented architecture that makes the implementation of new computational technologies very effective. 

\begin{figure*}[h]
\centering
\includegraphics[width=0.8\textwidth]{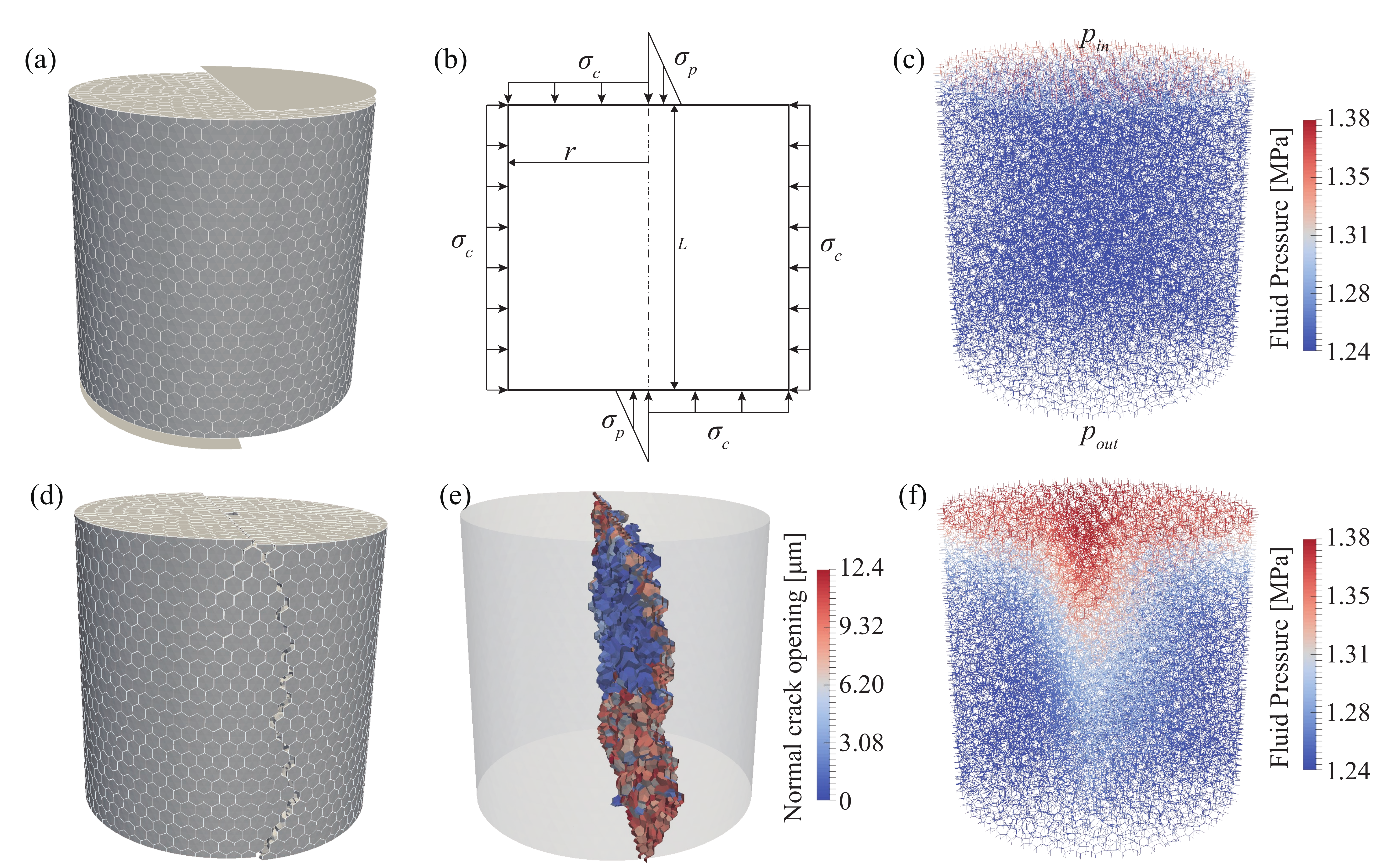}
\caption{(a) A cylindrical specimen with two opposing semi-circular disks modeled by LDPM, (b) a schematic diagram of the specimen and the loading configuration in a triaxial direct shear test, (c) a network of the flow elements occupying the computational domain and the fluid pressure boundary conditions. (d) A typical failure pattern, (e) crack opening profile, and (f) fluid pressure distribution for a typical failed specimen. }
\label{fig:Setup}     
\end{figure*}

\section{Numerical simulations of triaxial direct shear tests}
The M-LDPM framework is applied here to the numerical analysis of shale fracture permeability.
%by simulating triaxial direct shear tests conducted on carbonate-rich Marcellus shale \citep{frash2016high}. 
\subsection{Simulation setup}\label{sec:setup}
The simulated test setup is the one of recent triaxial direct shear tests performed at Los Alamos National Laboratory (LANL) to study shale fracture permeability behaviors at \textit{in situ} stress conditions and considering various effects such as confining pressure, \textit{in situ} stress, time-dependent factors, and stress cycling \citep{carey2015fracture,frash2016high,frash2017caprock}. 
The newly improved triaxial direct shear apparatus developed by \cite{frash2017caprock} provides better stress control by limiting specimen rotation in direct shear loading, and thus improves the accuracy and sensitivity of permeability measurements. 
%The simulations are performed by following the direct shear configuration. The setup is described as follows. 

A cylindrical specimen with length $H$, diameter $D$, and a 1:1 $H$ to $D$ ratio is placed between two opposing semi-circular disks, as depicted in Fig. \ref{fig:Setup}a. The disks are stiff enough so that can be simulated as rigid bodies with only vertical displacement allowed and are in direct contact with the upper-half and lower-half surface of the specimen. The load is applied through the upper disk, while the lower one is fixed. The specific loading configuration creates a plane of direct shear extending vertically through the specimen.
%, mimicking asymmetric loading generated by two opposed thrust collars in the experiments. 
In addition, a confining pressure is applied laterally on the external surface of the specimens and on the two upper and lower half surfaces not in contact with the loading platens. 
%and also on the half of each end of the specimen without the shear collar. 
The resultant boundary stresses are qualitatively illustrated in Fig. \ref{fig:Setup}b as per the description provided by \cite{frash2017caprock}.
%, is designed to reproduce the loading configuration provided by the triaxial direct shear apparatus. 
As one can see there is stress concentration, $\sigma_p$, near the direct shear plane and part of the upper and lower surfaces are not loaded. This is due to the fact that the specimens ends tend to rotate under load and they detach partially from the loading platens.

In the simulations, this is obtained by using a unilateral, frictional penalty contact model applied between the rigid disks and the specimen. The penalty contact model prevents compenetration of the specimen and platens but allows specimen to platen detachment. Hence, it is able to reproduce the experimental conditions and to replicate the stress state illustrated in Fig. \ref{fig:Setup}b. The penalty contact algorithm also includes simulation of Coulomb frictional effects. The static and kinetic coefficients of friction used in the simulations are 0.2 and 0.15, respectively. 
%
%Weixin:  did you use friction for the penalty model? what parameters did you use? if yes include 
%Cusatis: yes. I added the parameters

%
%in the simulations, while it was enabled by the stiff thrust collar that limits specimen rotation in the experiments \textcolor{blue}{The penalty approach needs more explanation}. 
%
%
\iffalse
%Weixin: I include the parameters in the text, let's remove the table
%Cusatis: OK. I checked the values in the text
\begin{table*}[h!]
  \centering
  \caption{Model parameters used for the triaxial direct shear simulations. }
    \begin{tabular}{llc}
    \hline
    Symbol [units] & Description & \multicolumn{1}{l}{Value} \\
    \hline
    $E_N$ [MPa] & Normal modulus & 118600 \\
    $\alpha$ [-] & Shear-normal coupling paramater & 0.1322 \\
    $\sigma_t$ [MPa] & Tensile strength  & 12.0 \\
    $\ell_t$ [mm] & Characteristic length & 1.0 \\
    $\sigma_s/\sigma_t$ [-] & Ratio between shear and tensile strength & 2.0 \\
    $\mu_0$ [-] & Internal friction coefficient & 0.18 \\
    $M_b$ [MPa] & Biot modulus & 61670 \\
    $\kappa_0$ [m$^2$] & Initial permeability & $1.0\times10^{-20}$ \\
    $\mu_w$ [Pa] & Dynamic viscosity of water at room temperature & $8.9\times10^{-4}$\\
    \hline
    \end{tabular}%
  \label{tab:Parameter}%
\end{table*}%
\fi
%
In the experiments, water was pumped onto the entire end face of the specimen for permeability measurement. In the simulations, inlet and outlet pressures, $p_{in}$ and $p_{out}$, respectively, were imposed on the nodal variables of the top and bottom boundary layers of the flow lattice model to create a differential pressure, $\Delta p = p_{in} - p_{out}$, across the core, as shown in Fig. \ref{fig:Setup}c. The corresponding upstream and downstream volumetric water flow rates were recorded continually during the entire test run. The apparent macroscopic permeability across the core area was calculated using Darcy's law applied to the entire sample. Due to dynamic processes occurring during the experiments (deformation, fines migration, etc.), permeability values changed slowly. The reported values represent pseudo-steady state where upstream and downstream flow rates were nearly equal \citep[see][]{frash2016high,frash2017caprock}.

%Various loading conditions and setups were explored. 
Three different effective confining pressures, $\sigma'_c = $ 4.4, 9.3, and 14.5 MPa, were considered in the experiments and simulations. Similar to the experiments, the effective confining pressure was calculated using the effective stress concept, i.e. $\sigma'_c = \sigma_c - 0.5b(p_{in} + p_{out})$, where $\sigma_c$ is the actual confining pressure applied on the specimen surfaces. A constant value of 0.6 was used for the Biot coefficient $b$ in this work. 
% Weixin: you earlier said that we used 0.6 ... we need to be consistent
% Cusatis: should be 0.6. Corrected.
%
%Discussion about the influence of Biot coefficient will be presented in Section \ref{sec:bioteffect}, and one can find that only the value of effective confining pressure matters. 
%
%Furthermore, geometrically similar specimens with increasing sizes were simulated in order to study the size effect on mechanical response and permeability measurement. The radius of the cylindrical specimens, $r$, is considered as a characteristic size. Four sizes, $r = 0.25, 0.4, 0.6$, and $0.8$ mm were considered while the geometrical ratios in all three dimensions were kept constant. 
%
%The effect of random microstructure is also explored by performing simulations on three specimens characterized by different placement of grains for each size.

The simulations were performed following the experimental sequence provided by \cite{frash2017caprock}. Firstly, the confining pressure was gradually increased to a target value while a minimal positive direct shear stress (up to 1.0 MPa) was maintained. Secondly, the axial displacement applied to the top semi-circular disk was steadily increased while maintaining constant confining pressure until major fracture creation was identified. Finally, loading ceased when the axial stress reached a plateau (residual stress). 

\begin{figure*}[h!]
\centering
\includegraphics[width=0.7\textwidth]{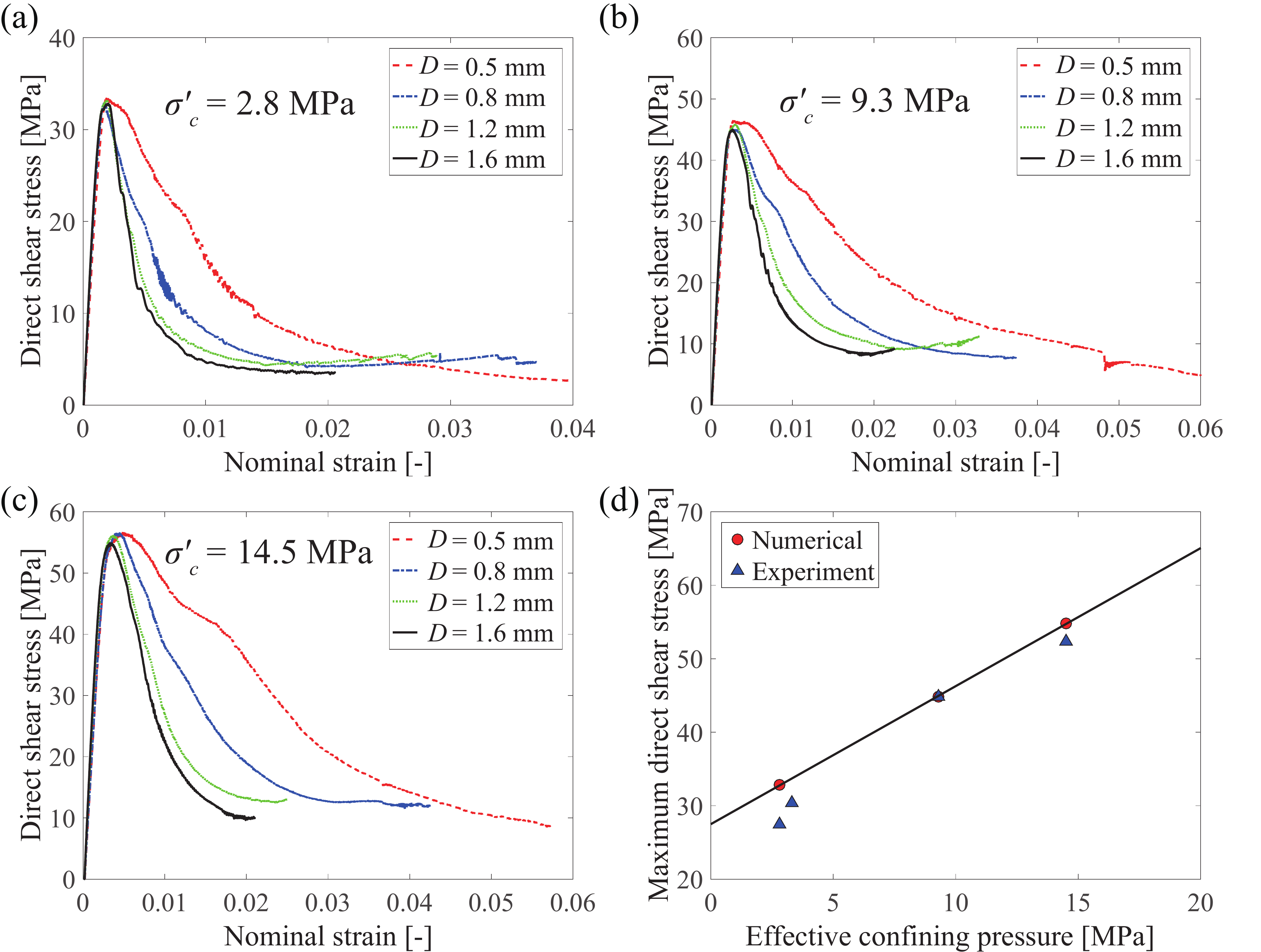}
\caption{ Direct shear stress versus normal strain curve obtained from triaxial direct shear simulations preformed on geometrically similar specimens with increasing size under effective confining pressures of (a) 2.8 MPa, (b) 9.3 MPa, and (c) 14.5 MPa. (d) Comparison of numerically calculated maximum direct shear stresses and the experimental data under various effective confining pressures. }
\label{fig:SS}     
\end{figure*}

\begin{figure*}
\centering
\includegraphics[width=1\textwidth]{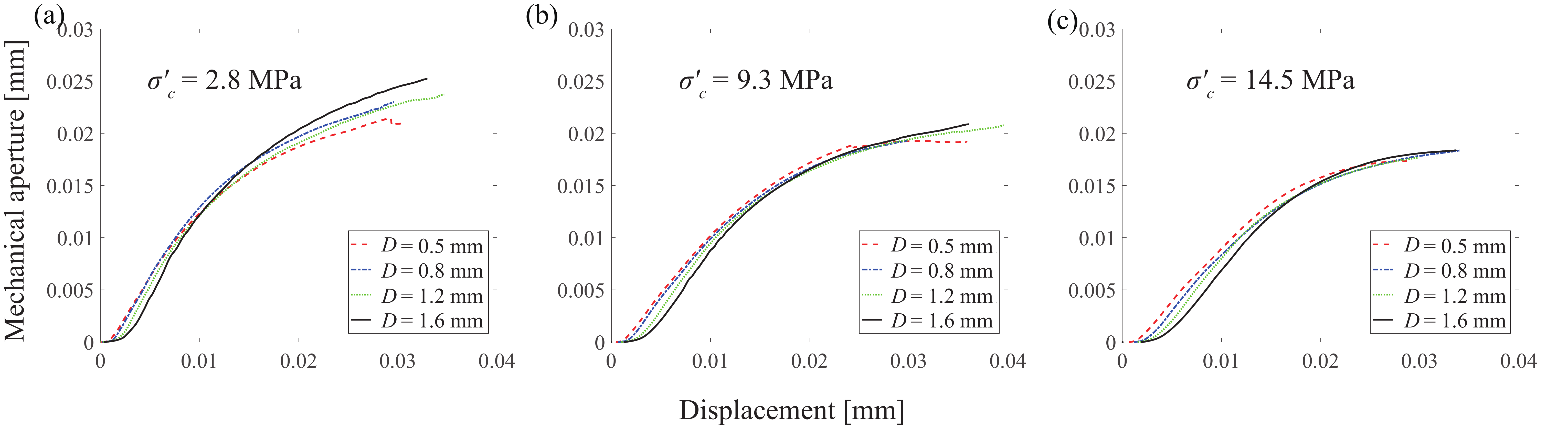}
\caption{ Variation of numerically calculated mechanical aperture with loading platen displacement under effective confining pressures of (a) 2.8 MPa, (b) 9.3 MPa, and (c) 14.5 MPa. }
\label{fig:AD}     
\end{figure*}

\subsection{Mechanical behavior}
The first step in the simulations is related to the LDPM parameter calibration. 

The generation of the shale mesostructure depends on the minimum and maximum size of the support particles (see Sec. \ref{sec:duallattice}), which, without specific information on the shale under study, were estimated \citep{li2017multiscale} as $d_0=$\SI{15}{\um} and $d_a=$\SI{35}{\um}. These values along with $v_a = 0.55$ lead to a LDPM cell size distribution characterized by minimum, average, and maximum cell size equal to 15, 30, and \SI{51}{\um}, respectively.
%
%Weixin: are these size for the spherical support particles or the cells?
%Cusatis: These size are for the cells. For particles, d_0 = 15 and d_a = 35
%Weixin: report the other parameters governing the mesostructure
%Cusatis：done

In addition, the  shale under study was reported to be statistically isotropic \citep{frash2017caprock} 
 since no anisotropy, pre-exiting flaws or bedding planes were observed in the tested specimens.

%As described previously, a set of geometric-related parameters was used to generate a granular lattice system, which requires observation and measurement of material internal structure and an investigation of its grain size distribution. Because of the lack of such information for the material under study, the geometric-related parameters reported in \cite{li2017multiscale} were used in this work, with a grain size distribution having a maximum, minimum, and mean diameter of 51, 13, and \SI{30}{\um}, respectively. The other parameters governing material behavior were calibrated to match the experimental results. 
%In addition, the intact material was assumed to be isotropic featured by a statistically isotropic random mesostructure because no anisotropy or pre-exiting flaws were found for the selected specimens as reported in the experiments. 

The experimental data reports an elastic modulus of $74 \pm 6$ GPa and a Poisson ratio of $0.30 \pm 0.02$. Since the samples were saturated with 
%Weixin: what is DI?
%Cusatis: DI water was mentioned in the experimental data, which means Deionized water
water and the tests were completed in a relatively short time period, the reported elastic properties must be considered to be undrained values. However, since the LDPM parameters govern the mesoscale mechanical behaviors of the solid phase, they must be related to the macroscopic properties measured under drained condition. Given a reasonable assumption of Biot and Skempton coefficients equal to 0.6
% Weixin: is 0.6 for both both Biot and Skempton coefficient?
% Cusatis: Yes. We used 0.6 for both
for typical shale \citep{suarez2013evaluating}, the drained elastic properties can be obtained by using the classic theory of poroelasticity \citep{detournay1993fundamentals,wang2000theory}. One has $68.8$ GPa and $0.21$ for the drained Young's modulus and the drained Poisson's ratio, respectively. 
% Weixin: incude here the undrained values
% CUsatis: added
With these values, the LDPM normal modulus, $E_N=118,600$ MPa, and shear-normal coupling parameter, $\alpha=0.13$, can be estimated with the micro-to-macro formulas reported by \cite{cusatis2011lattice}. The calculation was verified by simulating an undrained uniaxial compression test: the obtained undrained macroscopic Young's modulus and undrained macroscopic Poisson's ratio are 71 GPa and 0.29, respectively, which are in a good agreement with the experimental data. 
The LDPM tensile strength, $\sigma_t=12$ MPa, was estimated based on previous work by the authors \citep{li2017multiscale}.
%Weixin: how did you estimate \signma_t
%Cusatis: we don't have data to calibrate \sigma_t. The order of magnitude agrees with the LDPM shale paper 
%
The characteristic length, $\ell_t=1$ mm, was estimated \citep{cusatis2009cohesive}
%include here reference to my SE paper with cohesive law
according to measurements of the effective Fracture Process Zone (FPZ) length for Marcellus shale \citep{li2017characterization}. 

Finally, the shear strength, $\sigma_s=24$ MPa, and the internal friction coefficient, $\mu_0=0.18$, were calibrated by simulating the direct shear tests with various level of confinement. In order to save computational time, simulations were performed on reduced size samples. This was possible because under this loading condition the peak stress is basically size-independent. In order to demostrate this statement, specimens geometrically similar to the experimental ones but varting sizes were simulated: they had diamater equal to $D = 0.5, 0.8, 1.2$, and $1.6$ mm. 

Numerical results of the stress-strain curves for the geometrically similar specimens with increasing size are shown in Fig. \ref{fig:SS}a, b and c for various confining pressures. In these plots, similarly to \cite{frash2017caprock}, the net direct-shear stress acting on the direct-shear plane is calculated as $\tau_{DS}=(F_P - F_c)/(DH)$
%\begin{equation}
%\tau_{DS} = \frac{F_P - F_c}{DL}
%\end{equation}
where $F_P$ is the force applied on the rigid disk, and $F_c$ is the resultant force of the confining pressure applied on half of the specimen cross section. Also, the nominal strain is calculated as $\gamma=u/D$, where $u$ is the displacement of the upper loading platen. 

For each confining pressure, the curves for different sizes coincide in the elastic and nonlinear regime before reaching the peak. The peak stress is only minimally influenced by the sample size. The post-peak regime, however, is markedly different and the post-peak stress vs. strain curve tends to be steeper as the specimen size increases. This behavior is typical of quasi-brittle materials and can be explained by the size effect induced by damage localization and fracture \citep{bavzant1984size,bazant1997fracture}.
Indeed, as $\tau_{DS}$ approaches the peak, coalescence of microcracking occurs and finally leads to a localized fracture along the direct shear plane as one can see in Fig. \ref{fig:Setup}d and \ref{fig:Setup}e for a typical numerical response. 

%
% but diverge in the post-peak regime. Because the simulated specimens are tiny compared to the experiments, stable post-peak responses were captured in the simulations \textcolor{blue}{Why is this?}. One may note that the slope of the post-peak responses tends to be steeper as the specimen size increases. This phenomenon is due to
%the quasi-brittleness of rock and can be explained by the fracture size effect \citep{bavzant1984size}. The post-peak behaviors and their variation with specimen size are strongly related to the material's resistance to fracture, which is governed by the mesoscale fracture energy $G_t$ or the characteristic length $\ell_t$ in the LDPM framework. A meaningful scaling characteristic for the mechanical responses of the simulated specimens was assured by assigning a reasonable value of $G_t$ or $\ell_t$.  In this work, $\ell_t$ is assigned 1 mm according to the measurements of the effective Fracture Process Zone (FPZ) length for Marcellus shale \citep{li2017characterization}. 
%
If one extrapolates the numerical scaling results, snap-back at peak load with a loss of stability even under displacement control is expected for sufficiently large specimens. This indeed was observed in the experiments where the samples featured a catastrophic failure with a sudden drop of the applied load after attaining the maximum value \citep{frash2017caprock}. In addition, one can see in Fig. \ref{fig:SS}a, b, and c that, similarly to typical experimental data, the specimens under higher effective confining pressure exhibit a more ductile response and higher residual stress after failure. 

Figure \ref{fig:SS}d shows the comparison between the experimentally recorded and numerically calculated peak stresses vs. effective confining pressure. The numerical values are relevant to the simulations of samples with $D=0.5$ mm.
%Weixin: inlude value for D above
%Cusatis：added

The macroscopic internal friction angle, $\phi_{MC}$, and cohesion, $c_{MC}$, of the simulated material, were calculated from the maximum $\tau_{DS}$ values according to the classic Mohr-Coulomb criterion: $\phi_{MC} = 62^\circ$ and $c_{MC} = 27$ MPa. These values are in a good agreement with the experimental values $\phi_{MC} = 58\pm7^\circ$, and $c_{MC} = 26\pm2$ MPa reported in \cite{frash2017caprock}. The LDPM parameters were calibrated to slightly overestimate the experimental data in order to compensate for the size-effect discussed above.

The tested and simulated samples feature lateral expansion as a result of shear-dilation behavior. This can be characterized by the so-called mechanical aperture calculated by the difference of the lateral displacements between two diametrically opposite points located at half height of the specimens. The mechanical aperture results are shown in Fig. \ref{fig:AD}. It is clear that the mechanical aperture depends significantly on the confining pressure but its evolution versus the loading platen displacement is only minimally affected by the specimen size. The latter observation is key for the size scaling of the fracture permeability that will be discussed in the next section.

\begin{figure}
\centering
\vspace{-50pt}
\captionsetup[subfloat]{farskip=-110pt,captionskip=-60pt}
\subfloat[]{\includegraphics[width=0.45\textwidth]{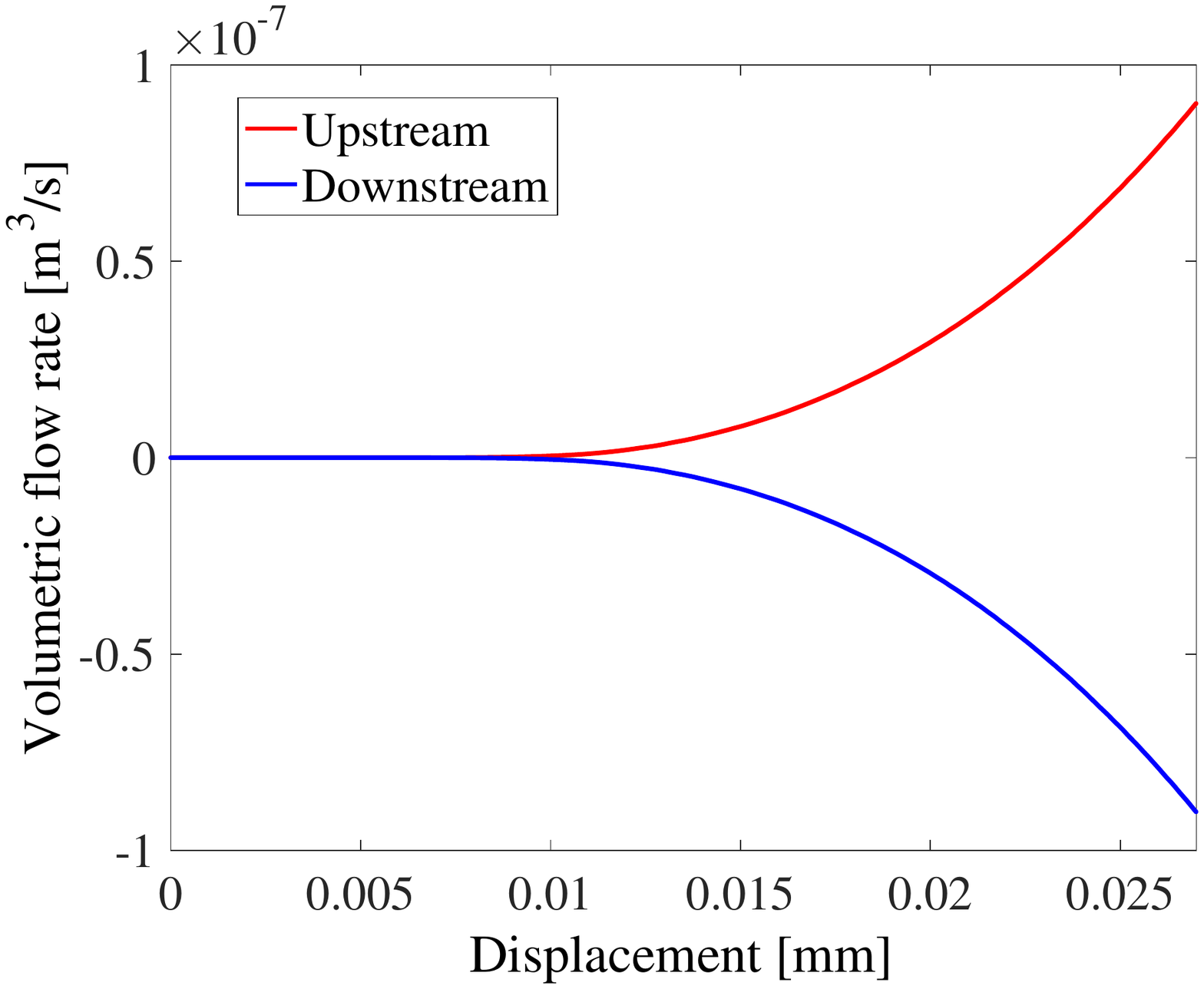}}
\subfloat[]{\includegraphics[width=0.45\textwidth]{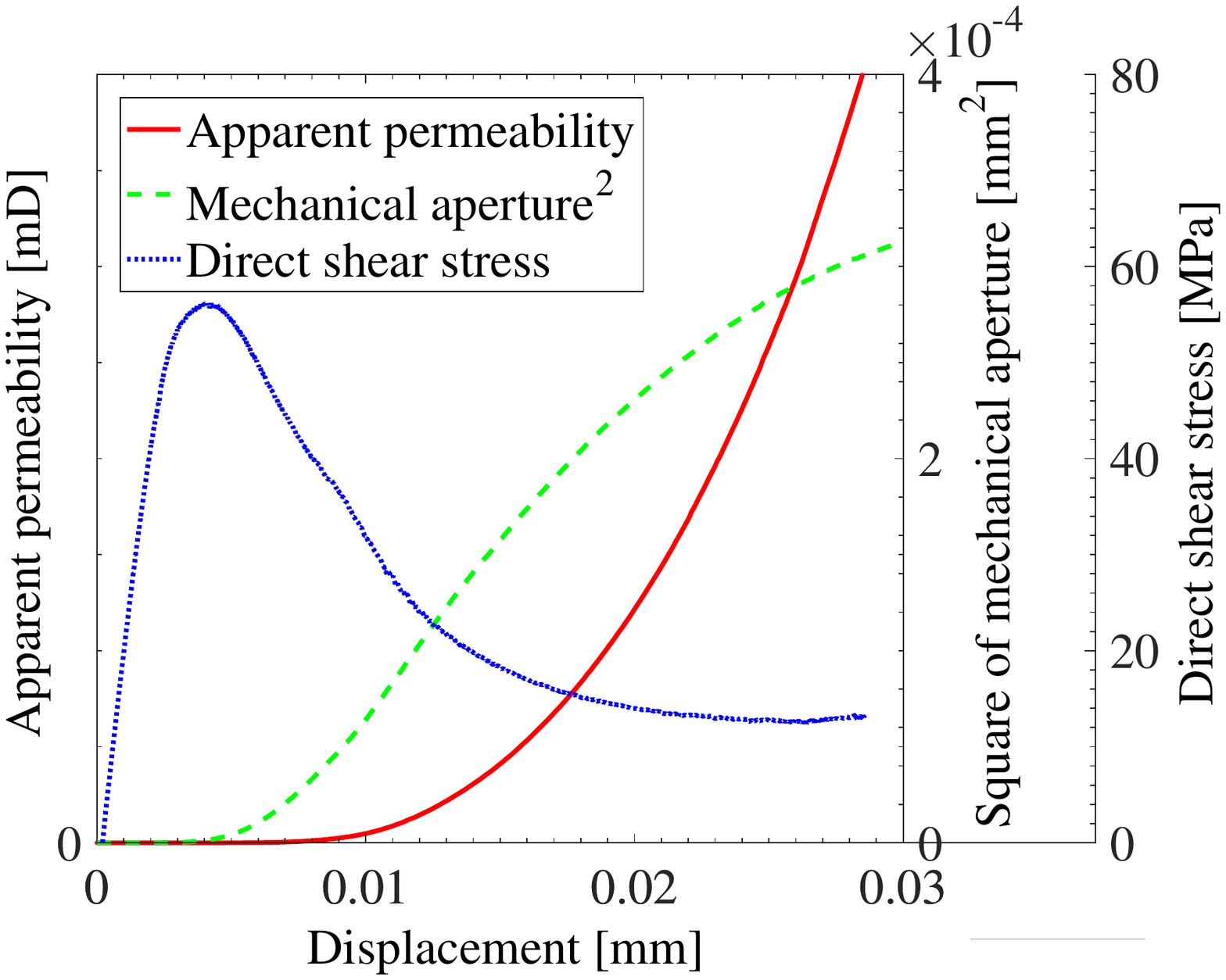}}
\vspace{-50pt}
\caption{(a) Typical upstream and downstream volumetric flow rate as a function of loading platen displacement, (b) typical permeability-aperture$^2$-stress relation as a function of loading platen displacement. }
\label{fig:kSD}     
\end{figure}

\subsection{Fracture permeability behavior}
The numerical predictions of the permeability measurements for the simulated specimens during the triaxial direct shear tests are presented in this section.

The dynamic viscosity of water, $\mu_w = 8.9\times10^{-4}$ Pa, the water bulk modulus, $K_w = 2.15$ GPa, and the water reference density, $\rho_{w0} = 1000$ kg/m$^3$, are values at room temperature and atmospheric pressure.  
In addition, the flow model depends on the intrinsic permeability and the Biot modulus. The value $\kappa_0=10^{-20}$ m$^2$=$10.13$ nD was estimated from the typical permeability of intact shale samples measured in laboratory \citep{soeder1988porosity}, and $M_b=61,670$ MPa was estimated from the experimentally reported elastic properties by using the classic theory of poroelasticity \citep{detournay1993fundamentals,wang2000theory}. 
%Weixin: check numbers above
%Weixin: how did you estimate these parameters?
%Cusatis, I added some words above

\begin{figure*}
\centering
\includegraphics[width=1\textwidth]{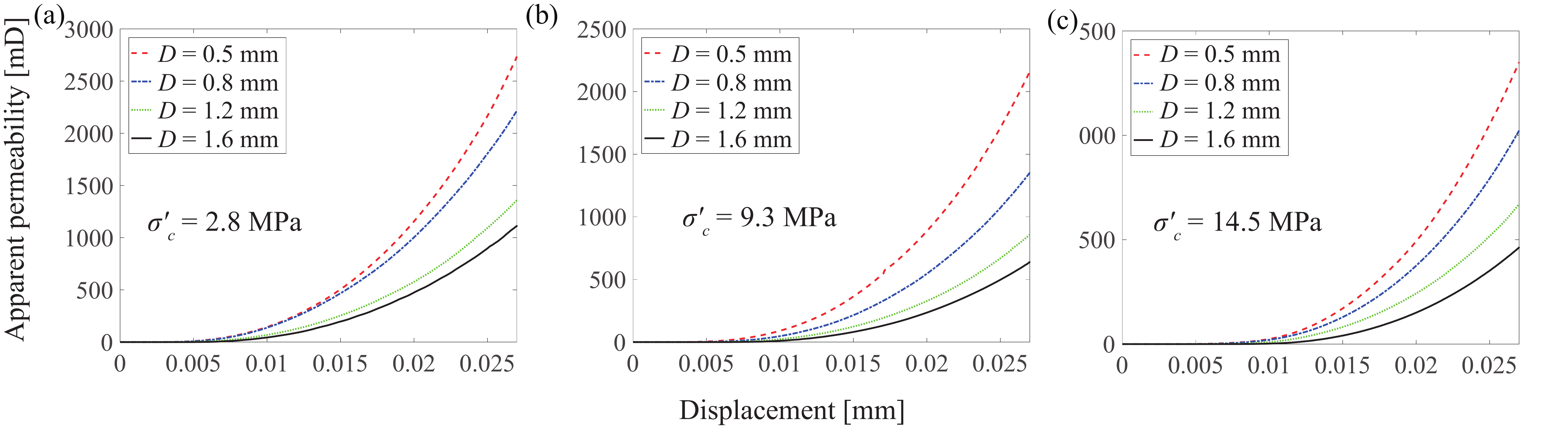}
\caption{ Numerically calculated apparent permeability as a function of loading platen displacement for the simulated triaxial direct shear tests under effective confining pressures equal to (a) 2.8 MPa, (b) 9.3 MPa, and (c) 14.5 MPa. Note the change in the y-axis scale across the three figures.}
\label{fig:kD}     
\end{figure*}

The typical upstream and downstream volumetric flow rate (discharge) data collected during one generic simulation run are plotted against the loading platen displacement in Fig. \ref{fig:kSD}a, and are used to calculate the macroscopic apparent permeability  through Darcy's law as follows
\begin{equation}
\label{darcy-exp}
K_{app} = \frac{4Q\mu_w H}{\pi D^2 (p_{in} - p_{out})} 
\end{equation}
where $Q$ is the absolute value of the downstream flow rate (downstream = -upstream when attaining steady state). Figure \ref{fig:kSD}b shows the typical permeability calculated using  Eq. \ref{darcy-exp} (solid line) from the numerical results as a function of the loading platen displacement. 
%Weixin: please remove colors from the picture 
%will do
For comparison, the same figure reports curves relevant to the square of the mechanical aperture (dashed line) and the direct shear stress (dotted line). The results in Fig. \ref{fig:kSD} are relevant to a sample with $D=1.2$ mm. 

In the elastic regime and before the peak the apparent permeability is very small (in the order of 10 nD). The mechanical aperture is negligible as well. 
%Weixin: this number should be in the order of $\kappa_0$, approximately 10 nD
%Cusatis: corrected
%
%is reported in the pre-peak regimie of the direct shear stress, and in the same regime, a slow rise of the mechanical aperture with displacement is mostly due to elastic deformation of the specimen. 
As the direct shear stress reaches the peak, a rapid rise of the mechanical aperture is observed, followed by a substantial increase of permeability. This is strongly associated with the development of localized fractures along the shear plane, as illustrated in Fig. \ref{fig:Setup}e. The localized fractures offer wide open channels permitting fluid flow. Correspondingly, the main flow is concentrated within the region of fracture localization, as shown in Fig. \ref{fig:Setup}f. 

The simulated permeability as well as the mechanical aperture keeps increasing gradually in the post-peak softening regime. This behavior is, of course, different from what observed in the experiments due to size effect. As explained earlier, the experiments feature a sudden drop of the stress as opposed to a gradual post-peak. As a result, in the experiments, the permeability spiked sharply upward as soon as failure occurred. This difference must be taken into account when comparing the computed fracture permeability values with the experimental ones.

In the presence of a single fracture, the fracture permeability is usually considered to be proportional to the square of mechanical aperture according to Poiseuille's formula. As shown in Fig. \ref{fig:kSD}b, the variations of the apparent permeability and the square of the measured mechanical aperture do not match. This indicates that global geometrical measurements, such as mechanical aperture, and fracture permeability are not closely correlated. In order to provide a sufficiently accurate prediction, a fine scale coupled model, such as M-LDPM presented in this work, is necessary to track the evolution of permeability as a consequence of interaction between local failure events and flow in three-dimension. 
%After attaining the residual stress, the simulations ceased because the current model is unable to address a topological change of grain structures due to large shear displacement. 

Figure \ref{fig:kD} presents the numerical results of the apparent permeability as a function of the loading platen displacement for the simulated specimens of various sizes and with different effective confining pressures. If one compares the numerically calculated apparent permeability values at a certain level of displacement, it can be seen that for the specimens with the same size, the permeability values decrease with an increase of effective confining pressure $\sigma_c'$. The same trend was observed in the experiments and many other rock fracture permeability studies \citep{gutierrez2000stress,davy2007permeability,jobmann2010investigation,zhang2013laboratory,zhou2016experimental}.
%add here relevant references
This is consistent with the previously discussed trend of decreased mechanical aperture for increased effective confining pressure.

Furthermore, one may note that the numerical results of the apparent permeability also depend on specimen size. A comparison of permeability for the specimens with different sizes yet under a same level of confinement shows that permeability tends to decrease with an increase of specimen size. The size dependence of the permeability measurements is caused by the fact that the damaged zone is localized and does not scale proportionally to the sample size.

%can be attributed to the generation of localized fractures, which are highly permeable, and thus provides a possibility of predicting fracture permeability behavior of larger specimens by formulating a scaling law. 
%More discussions of the fracture permeability size effect and the formulation of the scaling law will be presented in the next section. 

\subsection{Apparent permeability scaling analysis and predictions}
Due to the size dependence of the apparent permeability, the comparison between numerical predictions and experimental results must be performed with reference to an appropriate scaling law.

A simple scaling law correlating the fracture permeability measurements obtained from different sized specimens can be formulated if one assumes that (1) the width of the permeable band in which fracture localizes, $h$, is independent of specimen size; (2) The average opening or aperture of the permeable band is also size independent. In a direct shear test, such a permeable fracture band is indeed a shear band, and the opening or aperture of the shear band results from the shear-dilation behavior. For granular materials such as soil, these two assumptions are widely accepted: the shear band width depends mainly on grain size, and the shear-dilation is related to material constitutive behavior.

\begin{figure*}[h]
\centering
\includegraphics[width=0.8\textwidth]{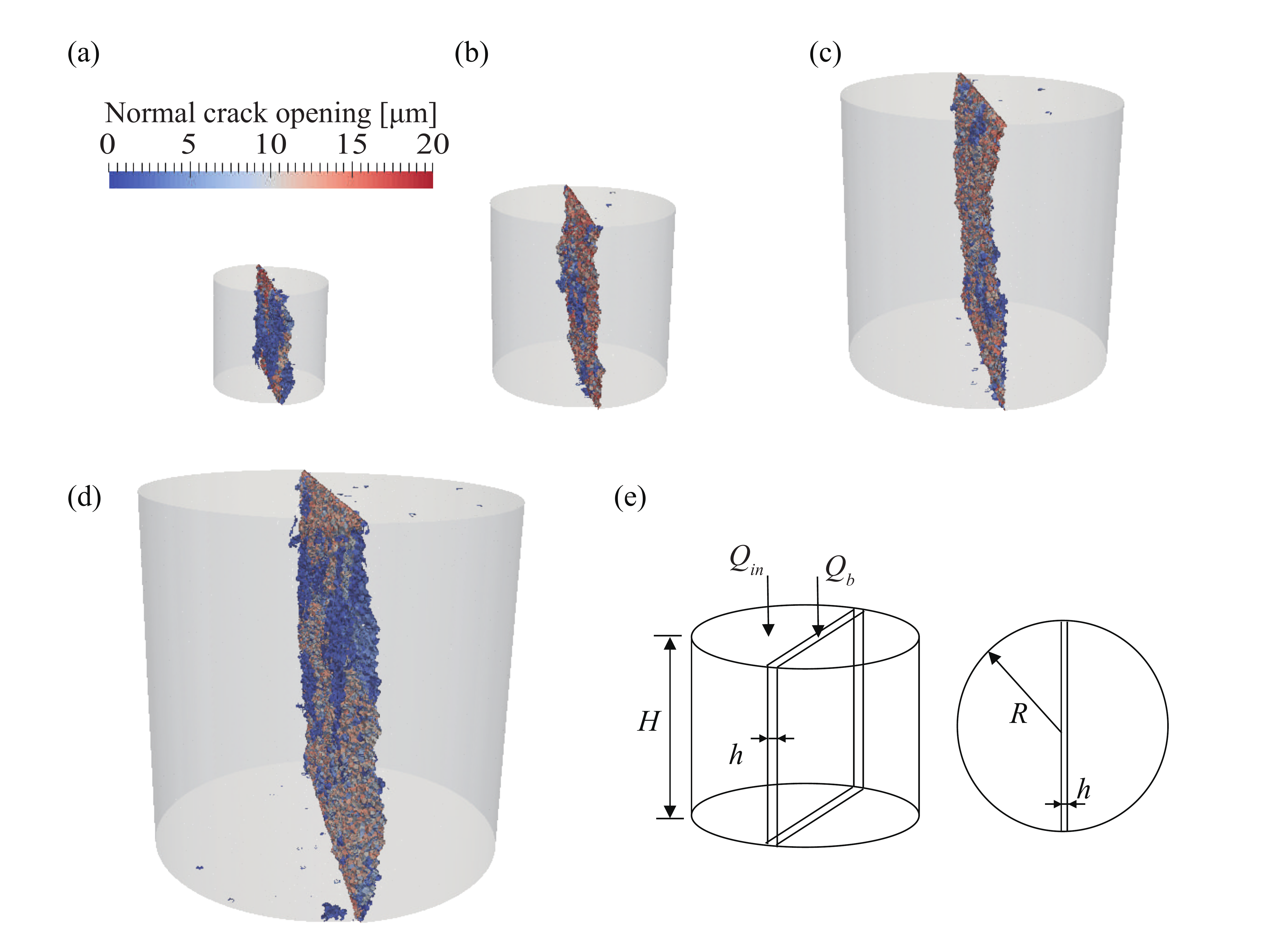}
\caption{ Normal crack opening profiles for the simulated geometrically similar specimens with various diameters: (a) $D = 0.5$ mm, (b) $D = 0.8$ mm, (c) $D = 1.2$ mm, and (d) $D = 1.6$ mm. (e) A schematic diagram of a parallel coupling between the flow rate through shear band $Q_b$ and the one through intact matrix $Q_{in}$. }
\label{fig:CrackSize}     
\end{figure*}

With reference to the simulation performed in the current study, the first assumption is supported by comparing the normal cracking profiles of the geometrically similar specimens with increasing size at the same level of displacement, as shown in Fig. \ref{fig:CrackSize}a, b, c and d. Note that the cracks are localized into a narrow band, i.e. shear band, with a statistically constant width for all investigated specimens. The second assumption is verified by the results on the mechanical aperture discussed in the previous section. 

%comparing the mechanical aperture for the simulated specimens with various sizes, as shown in Fig. \ref{fig:AD}. It can be seen that the variation of the aperture results converges to approximately the same curve at a relatively large level of displacement (post-peak regime) given a certain effective confining pressure regardless of specimen size. Therefore, these two assumptions hold for the simulated tests, which, on the other hand, can be attributed to the model's capability of capturing the grain-scale heterogeneity and various localization processes. 

\begin{figure}[h]
\vspace{-50pt}
\captionsetup[subfloat]{farskip=-110pt,captionskip=-60pt}
\centering
\begin{subfloat}[]{
\includegraphics[width=0.45\textwidth]{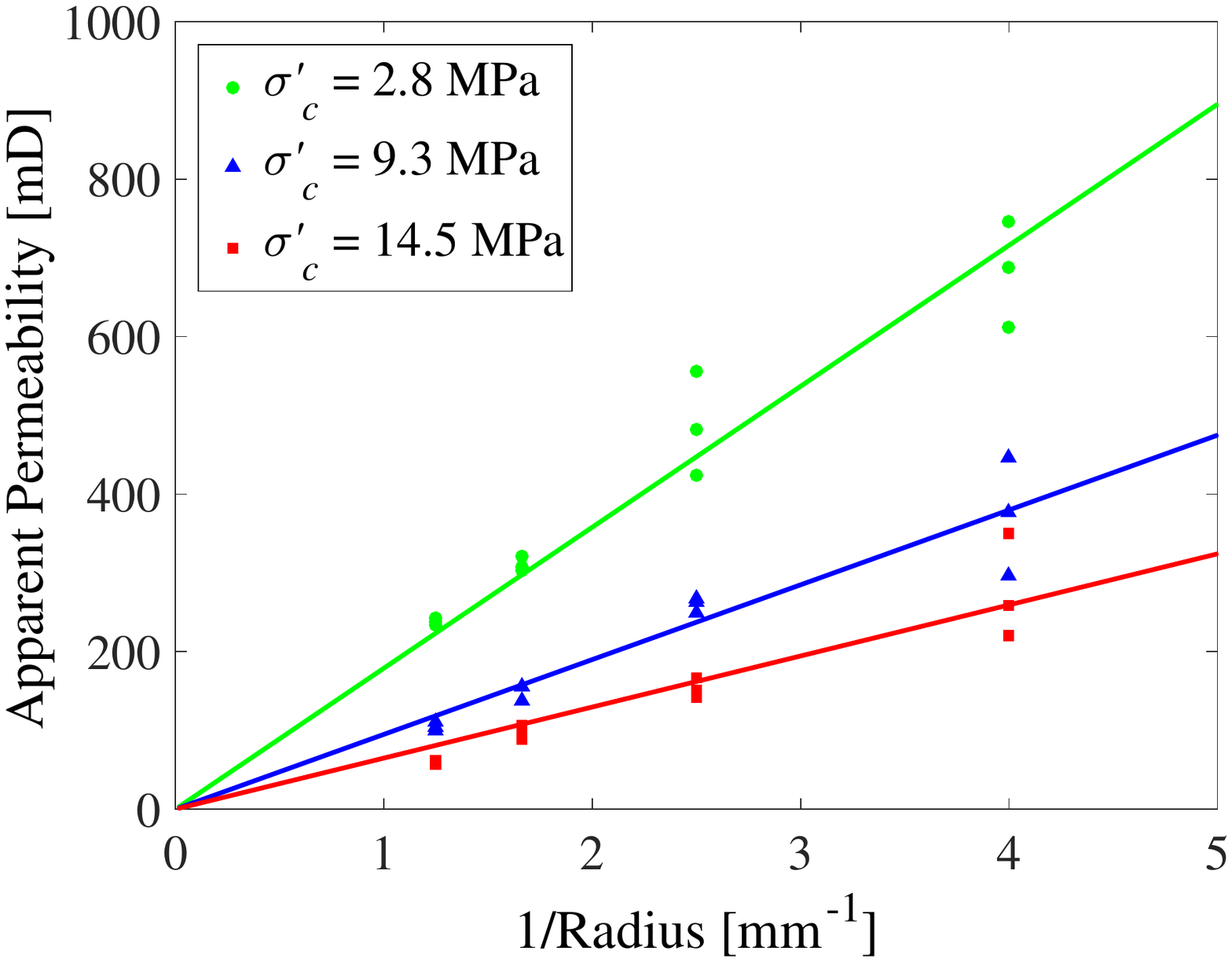}
}
%\vspace{-10\baselineskip}
\end{subfloat}
\begin{subfloat}[]{
\includegraphics[width=0.45\textwidth]{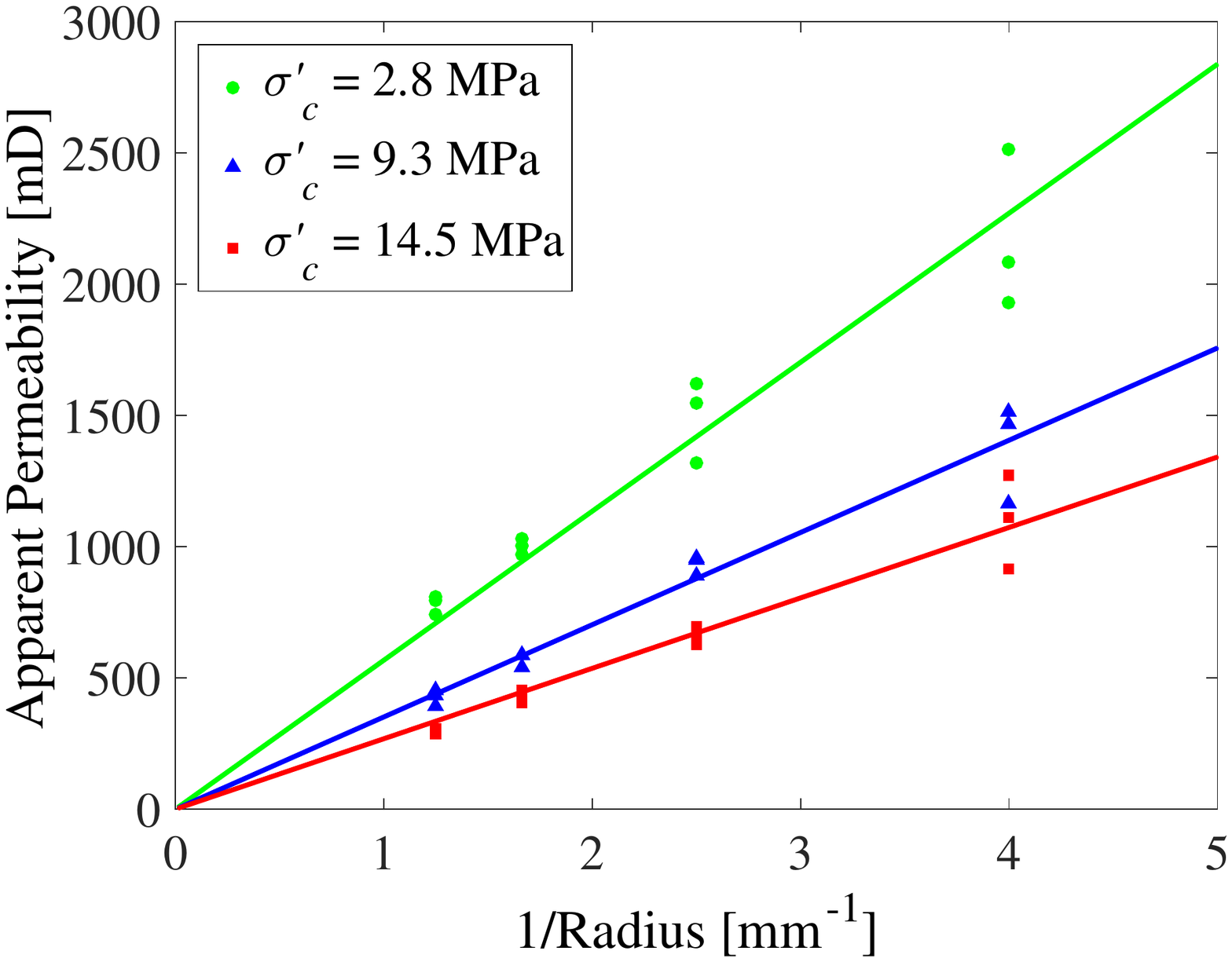}
}
\end{subfloat}
\vspace{-50pt}
\caption{ Apparent permeability sampled at displacement levels of (a) $u_1$ = 0.016 mm and (b) $u_2$ = 0.024 mm as a function of 1/radius of the specimens and linear regression analysis. }
\label{fig:Scaling}     
\end{figure}

Macroscopically, the flow across the specimen core can be decomposed into one component through the permeable shear band and the other one through the intact matrix. Correspondingly, the total flow rate $Q$ can be approximated by the sum of the flow rate through the shear band, $Q_b$, and the one through the intact matrix, $Q_{in}$, i.e. $Q = Q_b + Q_{in}$, where $Q = D_{app} \pi R^2 \Delta p/\mu_w H$, $Q_{in} = D_{in} \pi R^2 \Delta p/\mu_w H$, $Q_b = D_{b} 2R h \Delta p/\mu_w H$, $R=D/2$, and $D_{app}$, $D_{in}$, $D_{b}$ are the apparent macroscopic permeability, the macroscopic permeability of the intact material, the macroscopic permeability of the shear band, respectively. 
%This relationship is also shown schematically in Fig. \ref{fig:CrackSize}e, and can be reformulated as
This results in the following scaling law
\begin{equation}\label{eq:scaling-law-perm}
D_{app} =D_{in} + \frac{2h}{\pi R} D_b = D_{in} + B\frac{1}{R}
\end{equation}
The initial macroscopic permeability can be calculated approximately as $D_{in} \approx \kappa_0$; $D_{in}$ deviates from $\kappa_0$ due to the intrinsic heterogeneity of LDPM and the dual FLE system. However, since $D_{in}$ is calculated on a certain volume, the effect of the heterogeneity is filtered out for a large enough volume.
%The equation above relates the apparent permeability $\kappa_{app}$ with the specimen radius $r$, and can serve as a scaling law for the fracture permeability measurements from the direct shear tests given that $h$ and $\kappa_f$ are size independent, which are guaranteed by the first and second assumptions, respectively, as discussed above. 
Moreover, it worth pointing out that since $K_b$ is related to the mechanical aperture, it can be assumed not to change with size. Hence, since also the width of the permeable shear band $h$ is size independent, the scaling law in Eq. \ref{eq:scaling-law-perm} predicts a linear increase of the of the permeability with the inverse of the specimen radius. 

Let's consider now apparent permeability values sampled at two levels of displacement, $u_1 =$ 0.016 mm and $u_2$  = 0.024 mm. In both cases, the corresponding stresses are in the post-peak regime after fracture localization occurred. %These two values correspond to mechanical apertures equal to about ?? mm and ?? mm, respectively, irrespective of the the sample size. 
The corresponding mechanical apertures are about 0.013 mm and 0.017 mm for $\sigma'_c = 14.5$ MPa, which 
%This mechanical aperture 
are in the range of the experimentally measured mechanical aperture right after failure. 
%Weixin: includes values above 
%Cusatis: added

In order to take into account the effect of the randomness of the internal structure (captured by the proposed model), the apparent permeability values were calculated for three specimens with different internal structures randomly generated by the proposed model for each size and for each effective confining pressure. The results are plotted in Fig. \ref{fig:Scaling}a, and b for the two displacement levels, respectively, as function of the inverse of the specimen radius. A consistent trend of decreased permeability for increased specimen radius can be clearly identified. 

By performing a linear regression analysis of the generated numerical results, the constant $B$ in Eq. \ref{eq:scaling-law-perm} can be identified.
%Weixin: A should be basically equal to the intrinsic permeability \kappa_0 for all simulations. Did you get that? Which values did you get for A?
%Cusatis: The model is not accurate enough to get an accurate estimate of A. We actually prescribed A as \kappa_0 in the calculation. 
%
%Thanks to the formulation of the scaling law, it is now possible to predict the fracture permeability behavior of larger specimens through linear regression analysis. The regression equation can be written as $Y = a + bX$ with $Y = \kappa_{app}$, $X = 1/r$, $a = \kappa_0$, and $b = 2h \kappa_f/\pi$. The fitting of the numerical results based on the regression equation is also presented in Fig. \ref{fig:Scaling}. 

Once $B$ is identified, the scaling law in Eq. \ref{eq:scaling-law-perm} enables an extrapolation from the small size numerical simulations to the prediction of larger size specimens. Hence, considering the actual specimen radius of 12.7 mm in the experiments, one can predict the corresponding permeability values. The prediction results are listed in Table. \ref{tab:Kpredict}.

\begin{table}[h]
  \centering
  \caption{ Predicted apparent permeability }
    \begin{tabular}{m{1.1cm}>{\centering}m{2.0cm}>{\centering}m{2.0cm}>{\centering\arraybackslash}m{2.0cm}}
    \hline
    \multirow{2}[0]{1.4cm}{\scriptsize $\sigma'_c$, [MPa]} & \multicolumn{3}{c}{Apparent permeability, $D_{app}$, [mD]} \\
    \cline{2-4}
          & Simulation 1$^a$ & Simulation 2$^b$ & Experiments $^c$ \\
          \hline
    2.8   & 14.1  & 45.4  & 6.6 \\
    9.3   & 7.5   & 28.1  & 3.4 \\
    14.5  & 5.1   & 21.5  & 0.57 \\
    \hline
	\multicolumn{4}{l}{\scriptsize $^a$ Numerical results sampled at displacement level $u_1$ = 0.016 mm}\\
	\multicolumn{4}{l}{\scriptsize $^b$ Numerical results sampled at displacement level $u_2$ = 0.024 mm}\\
	\multicolumn{4}{l}{\scriptsize $^b$ Initial fracture permeability \citep{frash2017caprock} }    
    \end{tabular}%
  \label{tab:Kpredict}%
\end{table}%

Comparison of the predicted permeability values with the initial fracture permeability values measured in the experiments right after failure (see $4^{th}$ column of Table \ref{tab:Kpredict}) shows that the predicted values are of the same order of magnitude as the experimental data and follow the correct trend for increasing effective confining pressures. 
In general, the predicted values tend to be greater than the experimental values. This is probably due to the fact the formulated model simply formulates the meso-scale permeability to depend on the mesoscale normal crack opening but neglects the effect of the shear crack sliding which is known to reduce permeability \citep{klugesustainability}. 
%Weixin: include here refs to the french study showing this.
%Cusatis: added
Also, effects such as creep, chemical precipitation, particle mobilization, mylonitization and grain crushing induced during shearing, also play a role in the fracture permeability experiments and they are not included in the proposed model. However, a more sophisticated formulation requires specifically designed experiments which are not available at the moment.

\section{Conclusions}
A Multiphysics-Lattice Discrete Particle Model (M-LDPM) framework was formulated and used for the computational analysis of fracture permeability behavior of shale. This was obtained by simulating triaxial direct shear tests on Marcellus shale specimens of different sizes and under various confining pressures. Based on the obtained results, the following conclusions can be drawn. 

1) The proposed model can simulate the mechanical response of shale in the triaxial direct shear test, and successfully captured the poroelastic properties, the pressure-dependent strength properties, and the failure pattern. Benefiting from the model's capability of addressing grain-scale heterogeneity, the localized fracture and the size effect of the stress-strain responses can be captured accurately. 

2) The dual lattice topology enables the seamless coupling of the mechanical and flow behavior of the material at grain-scale. As a consequence, the variation of the apparent permeability induced by fracturing processes can be simulated by relating the hydraulic properties of flow elements with local cracking events. The numerical results demonstrated that a consistent trend of reduced fracture permeability for increased confining pressure can be captured properly by the proposed model in agreement with experimental evidence.

3) The scaling analysis of the permeability measurements obtained from the simulations performed on geometrically similar specimens enables a prediction of permeability for actual sized specimens by extrapolating the results of reduced scale specimens, which are required to make the computational cost of the simulations manageable. The predicted permeability values are in the same order of magnitude as the initial fracture permeability data measured in the experiments. 

%4) The numerical results obtained from the simulations with various Biot coefficients verified that the influence of changing Biot coefficient on the mechanical responses and fracture permeability measurements is incorporated in the definition of effective confining pressure in the triaxial direct shear tests. 

\section*{Acknowledgements}
The work of WL and GC was supported by the Center for Sustainable Engineering of Geological and Infrastructure Materials (SEGIM) and the Quest high performance computing facility at Northwestern University. The Work of XZ was supported with ES3 R\&D resources. 
The work of WC and LF was supported by the Department of Energy's Basic Energy Sciences program under grant DE-AC52-06NA25396/FWP-LANL20171450.

\bibliographystyle{spbasic}      % basic style, author-year citations
\bibliography{bib1}   % name your BibTeX data base


\end{document}